\documentclass[prb, twocolumn, superscriptaddress, floatfix]{revtex4-2}
\usepackage[english]{babel}

\usepackage{amsmath} 
\usepackage{amssymb} 
\usepackage{upgreek}
\usepackage{physics}
\usepackage{amsfonts}
\usepackage{graphicx} 
\usepackage{tikz}
\usepackage{hyperref} 

\usepackage{booktabs}

\newcommand{\uv}{Departament de Física Teòrica and IFIC, Universitat de València-CSIC, 46100 Burjassot (València), Spain}
\begin{document}

\title{Quantum Natural Gradient optimizer on noisy platforms: QAOA as a case study}
\author{Federico Dell'Anna}
\affiliation{Dipartimento di Fisica e Astronomia dell’Università di Bologna, I-40127 Bologna, Italy}
\affiliation{INFN, Sezione di Bologna, I-40127 Bologna, Italy}
\author{Rafael Gómez-Lurbe}
\affiliation{\uv}

\author{Armando Pérez}
\affiliation{\uv}
\author{Elisa Ercolessi}
\affiliation{Dipartimento di Fisica e Astronomia dell’Università di Bologna, I-40127 Bologna, Italy}
\affiliation{INFN, Sezione di Bologna, I-40127 Bologna, Italy}

\begin{abstract}
We investigate the performance of the Quantum Natural Gradient (QNG) optimizer in the presence of noise. Specifically, we evaluate the efficacy of QNG within the Quantum Approximate Optimization Algorithm (QAOA) for finding the ground state of the Transverse Field Ising Model (TFIM). Its performance is benchmarked against the Vanilla Gradient Descent optimizer across two prominent quantum computing platforms: Rydberg atoms and superconducting circuits. Our analysis includes simulations under both idealized noise-free conditions and realistic noisy environments based on calibration data from actual devices. Results demonstrate that QNG consistently outperforms Vanilla Gradient Descent, exhibiting faster convergence on average and greater robustness against random initializations of parameters. This robustness is attributed to the distance regularization in parameter space inherent to QNG. Additionally, QNG achieves a higher convergence rate to the solution, effectively avoiding certain local minima. These findings highlight QNG as a promising tool for optimizing variational quantum algorithms in noisy intermediate-scale quantum (NISQ) devices.
\end{abstract}
\maketitle
\section{Introduction}
Hybrid quantum-classical variational algorithms \cite{mcclean2016theory,peruzzo2014variational,moll2018quantum} are essential for exploring the potential of noisy intermediate-scale quantum (NISQ) devices  \cite{preskill2018quantum}. These algorithms combine classical computation with quantum processing, where classical resources are used to optimize the parameters of quantum states. By employing heuristic techniques, they tackle variational problems, for instance combinatorial optimization tasks, which are both prevalent and critical in many real-world applications \cite{de2018analysis}. Consequently, these methods have garnered substantial interest from industries looking to leverage quantum computing, as solving such problems remains notoriously difficult for classical algorithms alone.

The Quantum Approximate Optimization Algorithm (QAOA) \cite{farhi2014quantum} stands out among hybrid variational algorithms for its promise in achieving quantum speedups on NISQ devices, drawing considerable interest \cite{zhou2020quantum}. QAOA has been successfully implemented across a variety of experimental platforms, such as Rydberg atom arrays \cite{tibaldi2025analog, ebadi2022quantum}, superconducting processors \cite{harrigan2021quantum}, and trapped-ion systems \cite{pagano2020arinjoy}.

Similar to other hybrid algorithms, QAOA involves applying a series of parameterized quantum gates to a quantum state with the goal of minimizing the expectation value of an observable, typically the system's Hamiltonian. The classical part of the algorithm optimizes the gate parameters to achieve this minimization, requiring frequent evaluations of the Hamiltonian's expectation value \cite{huang2020predicting}. The interplay between quantum execution and classical optimization demands repeated runs of the quantum circuit, making the process resource-intensive. This implies that finding the minimum of the cost function in fewer steps plays a crucial role.

A significant challenge in this approach is the occurrence of barren plateaus: large flat areas of the parameters landscape with exponentially small gradients as the number of qubits and circuit depth increase \cite{mcclean2018barren}. This problem can be exacerbated by noise \cite{wang2021minimizing} or by employing cost functions dependent on global observables \cite{cerezo2021cost}.

This study will focus on a paradigmatic problem: finding the ground state preparation of the Transverse Field Ising Model (TFIM). In particular, we aim to combine the Quantum Approximate Optimization Algorithm (QAOA) with the Quantum Natural Gradient (QNG) optimizer and evaluate its performance on simulations of realistic noisy quantum devices.
The Quantum Natural Gradient was originally proposed in \cite{stokes2020quantum} to enhance the optimization of variational quantum algorithms. Several studies have since demonstrated the potential advantages of using QNG techniques in these algorithms \cite{wierichs2020avoiding,lopatnikova2021quantum,roy2023efficient}.

The main idea is to leverage information about the geometry of the Hilbert space. Specifically, this involves scaling the gradient of the cost function by the inverse of the Fubini-Study metric, which is the Riemannian metric in the space of parameterized quantum systems \cite{ercolessi2010equations}. This approach ensures that the algorithm follows the correct direction of steepest descent, thereby accelerating its convergence towards the solution.

Building on this foundation, further research extended the quantum natural gradient approach to tackle challenges such as noise and non-unitary evolutions in quantum systems \cite{koczor2022quantum}, using their own noise models along with some procedure to approximate the Fubini-Study metric.

In light of the existing work, our aim is to benchmark the performance of the QNG optimizer against the Vanilla Gradient Descent optimizer on two major quantum computing platforms under realistic noise models: Rydberg atoms and superconducting circuits.

In the current NISQ era, gradient-free optimization techniques are often preferred for QAOA due to their lower measurement overhead and simpler implementation. In this work, we deliberately focus on gradient-based methods---not as a recommendation for near-term optimization, but rather as a theoretical and exploratory benchmark to assess the robustness of QNG under realistic noise models.

Indeed, the primary objective of this work is not to demonstrate state-of-the-art optimization performance, but rather to investigate how the QNG responds to realistic noise conditions typical of current quantum platforms. For this reason, we adopt the TFIM as a toy model, serving as a well-understood and controllable testbed for benchmarking the behavior of QNG.
\bigskip

This work is organized as follows: In Sec. \ref{QAOA}, we briefly review the QAOA algorithm, the Transverse Field Ising Model, and the Quantum Natural Gradient optimizer. Sec. \ref{Rydberg} introduces the basics of the Rydberg atoms platform, the implementation of QAOA on it, the noise sources considered, and the numerical studies performed. The subsequent Sec. \ref{IBM} presents the quantum circuit implementation of QAOA, the realistic noise model used, and the simulations conducted on this platform. Finally, in Sec. \ref{Conclusions}, we summarize the conclusions drawn from this work, as well as outline future studies and open problems in this research line.

\section{Quantum Approximate Optimization Algorithm and Quantum Natural Gradient}\label{QAOA}

The Quantum Approximate Optimization Algorithm (QAOA) is a hybrid quantum-classical algorithm originally inspired by the adiabatic theorem and designed to solve combinatorial optimization problems. It was first proposed by Edward Farhi and Jeffrey Goldstone in 2014 \cite{farhi2014quantum} and has since become one of the most promising quantum algorithms for near-term quantum devices.
The algorithm consists of two main components: a classical optimizer and a quantum circuit that prepares a parameterized trial state on a quantum computer. The key idea is to iteratively adjust the parameters to minimize the expectation value of the cost Hamiltonian, thereby approaching the optimal solution.

\subsection{QAOA Circuit}\label{QAOACIRC}

\begin{figure*}
    \centering
    \includegraphics[width=0.8\linewidth]{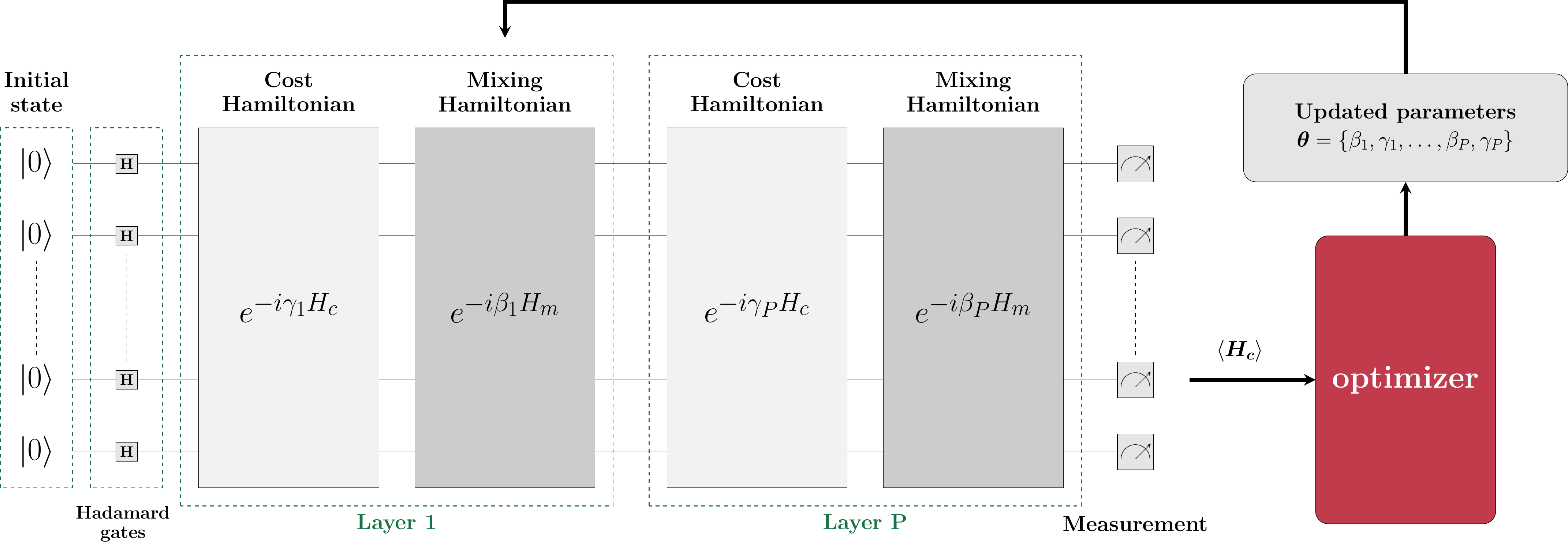}
 \caption{QAOA scheme: starting from an initial state, Hadamard gates prepare a superposition. Each layer alternates between the application of the Cost Hamiltonian \( e^{-i\gamma H_c} \) and the Mixing Hamiltonian \( e^{-i\beta H_m} \). The parameters \(\gamma\) and \(\beta\) are iteratively optimized to minimize the cost function \(\langle H_c \rangle\) after the measurement phase.}
    \label{fig:QAOA}
\end{figure*}

The QAOA circuit, as shown in Fig.~\ref{fig:QAOA}, involves the following steps:
\begin{enumerate}
    \item \textbf{Initialization}. The algorithm starts with an initial state: the uniform superposition state \(|+\rangle^{\otimes N}\), where \(N\) is the number of qubits.
    
    \item \textbf{Parameterized evolution}. The state is then evolved using two types of unitary operators:
    \begin{itemize}
        \item \textit{Cost Hamiltonian Unitary}: \(U_c(\gamma) = e^{-i \gamma H_c}\)
        \item \textit{Mixer Hamiltonian Unitary}: \(U_m(\beta) = e^{-i \beta H_m}\)
    \end{itemize}
    Here, \(H_c\) is the cost Hamiltonian and \(H_m\) is the mixer Hamiltonian. To explore the whole Hilbert space, it is necessary to have \( [H_c, H_m] \neq 0 \). Then, the mixer Hamiltonian is chosen as \( H_m = \sum_{i=1}^N \sigma^x_i \), where \( \sigma^x_i \) are Pauli-X operators acting on the \( i \)-th qubit.
    
    \item \textbf{Layered structure}. The evolution is applied in layers, where each layer consists of a pair of unitaries \(U_c(\gamma)\) and \(U_m(\beta)\). The depth of the QAOA circuit is determined by the number of layers \(P\), leading to a trial state: 
    
    $|\psi(\vec{\gamma}, \vec{\beta})\rangle = U_m(\beta_P) U_c(\gamma_P) \cdots U_m(\beta_1) U_c(\gamma_1) |+\rangle^{\otimes N}$ 
    
    Here, \(\vec{\gamma} = (\gamma_1, \gamma_2, \ldots, \gamma_P)\) and \(\vec{\beta} = (\beta_1, \beta_2, \ldots, \beta_P)\) are the parameters to be optimized.
    
    \item \textbf{Measurement and optimization}. The expectation value of the problem Hamiltonian is measured in the trial state:
    \begin{equation}
   E_c= \langle \psi(\vec{\gamma}, \vec{\beta}) | H_c | \psi(\vec{\gamma}, \vec{\beta}) \rangle   
    \end{equation}
    A classical optimizer adjusts the parameters \(\vec{\gamma}\) and \(\vec{\beta}\) to minimize this expectation value, iterating the process until convergence.
\end{enumerate}
The QAOA merges the advantages of both quantum and classical computing. 

\subsection{Application to the Transverse Field Ising Model}
\label{Application to TFIM}

In our study, we apply QAOA to the TFIM, a well-known model in quantum mechanics and statistical physics. The Hamiltonian for the 1D TFIM is given by:
\begin{equation}
H = -J \sum_{i=1}^{N} \sigma^z_i \sigma^z_{i+1} - h \sum_{i=1}^N \sigma^x_i
    \label{eq:H_Ising}
\end{equation}
where \(\sigma^z_i\) and \(\sigma^x_i\) are Pauli-Z and Pauli-X operators acting on the \(i\)-th qubit, \(J\) is the interaction strength between neighboring spins, and \(h\) is the transverse magnetic field strength.
By optimizing the parameters of the QAOA circuit, we aim at finding the ground state of this Hamiltonian. From now on, we work setting $J=1$, $h=0.5$ and with periodic boundary conditions.

As previously mentioned, the Quantum Approximate Optimization Algorithm (QAOA) follows a sequence of well-defined steps while allowing flexibility in the choice of Hamiltonians used to evolve the system. In particular, for the Transverse Field Ising Model (TFIM), we modify the standard structure of QAOA by defining the unitary operators governing the evolution as \( U_{zz}(\gamma) = e^{-i\gamma H_{zz}} \), where \( H_{zz} = -J \sum_{i=1}^{N} \sigma^z_i \sigma^z_{i+1} \), for the first evolution of the state. As the mixing Hamiltonian, we use the standard form \( U_{m}(\beta) = e^{-i\beta H_{m}} \), where \( H_m = -h \sum_{i=1}^{N} \sigma^x_i \). For our numerical simulations, we set \( J=1 \) and \( h=0.5 \). However, as the cost function, we measure the expectation value of the full Hamiltonian given in Eq.~\ref{eq:H_Ising}.  

This modification is motivated by the fact that the conventional QAOA structure typically employs a mixing Hamiltonian that has the same form as the second term in the problem Hamiltonian (Eq.~\ref{eq:H_Ising}). By adjusting the standard procedure, we eliminate this redundancy.

By implementing the algorithm using the QuTiP library~\cite{qutip2013}, we verified that the ground state is reached at a circuit depth of \( P = \left\lfloor \frac{N}{2} \right\rfloor \), by testing different values of \( P \)  using the quantum natural gradient as optimizer. This preliminary check will also be performed in the implementations based on Rydberg atoms and IBM superconducting qubits, using the Pulser and PennyLane backends,
respectively.

This result implies that circuits with such a depth are sufficiently expressive to capture the ground state while preserving the symmetries of the Hamiltonian. Hence, we adopt this parameterization for the algorithm's evolution, subject to validation during implementation on the selected platform, which inevitably introduces certain approximations, both in the presence and absence of noise.

The energy value obtained at the end of the optimization procedure $E_c$ can be compared with the exact one $E_0$, allowing us to compute the accuracy:
 \begin{equation}   
 \delta E := \frac{E_c - E_0}{\abs{E_0}}.  
 \label{eq:accuaracy}
\end{equation}
The exact energy of the TFIM with \( J=1 \) can be calculated analytically~\cite{wierichs2020avoiding,lieb1961two}:
\begin{equation}
E_0 = -E_1 - 2 \sum_{q=1}^{r} \frac{1}{1 + h^2 + 2h \cos \alpha_q}, 
\end{equation}
where $r=\left\lfloor N/2 \right\rfloor $ and
\begin{equation}
\begin{cases} 
\alpha_q := \frac{(2q - 1)\pi}{N},& \quad E_1 := 0; \qquad \quad for\; N \, even\\
\alpha_q := \frac{2q\pi}{N}, & \quad E_1 := 1+h; \quad for\; N \, odd
\end{cases}
\end{equation}
Our implementation of QAOA on real quantum platforms, such as Rydberg atoms and superconducting circuits, aims to demonstrate the algorithm's robustness and effectiveness, even in the presence of various sources of noise.

\subsection{Quantum Natural Gradient}

Vanilla gradient descent, or simply gradient descent, is an optimization algorithm used to minimize a function by iteratively moving in the direction of the steepest descent, defined as the negative gradient. It is widely used in machine learning and statistics for optimizing models by adjusting their parameters \cite{ruder2016overview}. 

Here's a brief overview of the method:
we start with an initial guess for the parameters, denoted as 
\( \boldsymbol{\theta} = (\theta_1, \theta_2, \ldots, \theta_n)\).
Next, we calculate the gradient of the loss function \(C( \boldsymbol{\theta})\) and update the parameters by moving in the opposite direction of the gradient. The update rule for each parameter \(\theta_i\) is given by:
\begin{equation}
\boldsymbol{\theta}_{t+1} = \boldsymbol{\theta}_{t} - \eta \nabla C( \boldsymbol{\theta}_{t}),       
\end{equation}
where \(\eta\) is the learning rate, a small positive number that controls the step size.
These steps are repeated until convergence that is typically defined by a condition such as:
\begin{equation}\label{conv}
\| C( \boldsymbol{\theta}_{t+1}) - C( \boldsymbol{\theta}_t) \| < \epsilon_{stop},       
\end{equation}
where \(\epsilon\) is a small threshold indicating that the changes in the parameters are sufficiently small.

The update rule for gradient descent can be derived from the following optimization problem \cite{stokes2020quantum}: 
\begin{equation}
    \boldsymbol{\theta}_{t+1} = \arg\min_{\boldsymbol{\theta} \in \mathbb{R}^d} \left[ \langle \boldsymbol{\theta} - \boldsymbol{\theta}_{t}, \nabla C(\boldsymbol{\theta}_{t}) \rangle + \frac{1}{2\eta} \|\boldsymbol{\theta} - \boldsymbol{\theta}_{t}\|^2_{2} \right]
\end{equation}

Examining this expression naturally raises the following question: we have assumed that the geometry of the parameter space is Euclidean, but what happens if this assumption does not hold and the space instead has a different Riemannian structure? This question was posed by the authors in  \cite{Amari98}, leading to the introduction of the Natural Gradient Algorithm. 

In this algorithm, the parameter update rule is modified by multiplying the gradient by the inverse of the Fisher Information Matrix, \( I^{-1}(\theta) \). The Fisher Information Matrix serves as the Riemannian metric that captures the geometry of the statistical model, guiding the gradient in the true direction of steepest descent. By accounting for the local curvature of the probability distribution space, the update rule becomes more natural and can potentially lead to faster optimization in the parameter space  \cite{amari2019fisher, amari1996neural,goodfellow2016regularization,amari2000adaptive}.

In direct analogy to the Natural Gradient Algorithm, the authors in \cite{stokes2020quantum} proposed extending this approach to the quantum setting. In this context, the metric tensor is defined on the manifold of parameterized quantum states. For pure states, this metric is known as the Fubini-Study metric \cite{Bengtsson_Zyczkowski_2006}:
\begin{align}
g_{ij}(\boldsymbol{\theta}) &= \text{Re} \Big[ 
\langle \partial_i \psi(\boldsymbol{\theta}) | \partial_j \psi(\boldsymbol{\theta}) \rangle \nonumber \\
&\quad - \langle \partial_i \psi(\boldsymbol{\theta}) | \psi(\boldsymbol{\theta}) \rangle 
\langle \psi(\boldsymbol{\theta}) | \partial_j \psi(\boldsymbol{\theta}) \rangle \Big],
\end{align}
where \( |\psi(\boldsymbol{\theta})\rangle \) is a parameterized quantum state, and we use the shorthand \( \partial_i \equiv \frac{\partial}{\partial \theta_i} \).

With this metric, the update rule for the Quantum Natural Gradient (QNG) is expressed as:
\begin{equation}
\boldsymbol{\theta}_{t+1}=\boldsymbol{\theta}_{t}-\eta g^{+}(\boldsymbol{\theta}_{t})\nabla C(\boldsymbol{\theta}_{t}),
\end{equation}
where $g^{+}(\boldsymbol{\theta}_{t})$ denotes the pseudo-inverse of the metric tensor, which is particularly useful when handling singular matrices.

We can further specify the update rule for the parameters within the QAOA ansatz:
\begin{equation}
\boldsymbol\theta_{t+1} = \boldsymbol\theta_{t} - \eta g^{+}(\boldsymbol\theta_{t}) \nabla E(\boldsymbol{\theta}_{t}),
\end{equation}
where \( \boldsymbol{\theta} = (\gamma_1, \beta_1, \gamma_2, \beta_2, \dots, \gamma_P, \beta_P) \). Here, $E(\boldsymbol{\theta}) = \mathrm{Tr}[\rho_{2P}(\boldsymbol{\theta}) H_c] $ represents the expectation value of the cost Hamiltonian.
Additionally, \( \rho_{2P}(\boldsymbol{\theta}) \) represents the state obtained at the end of the circuit, i.e., after evolving the system with \( 2P \) parameters. In general, we assume this state to be mixed.

Intuitively, the Fubini-Study metric measures the sensitivity of the quantum state to changes in the parameters. Specifically, the Fubini-Study matrix will exhibit larger entries when the parameter variations significantly impact the quantum state and smaller entries when the effect is minimal. Consequently, since the update rule scales the gradient by the inverse of the metric, the QNG method effectively regularizes such changes in the parameter space.

Interestingly, this metric is related to the Quantum Fisher Information Matrix (QFIM), which defines the ultimate precision limit for quantum multiparameter estimation through the quantum Cramér-Rao bound \cite{Helstrom, Holevo}. Specifically, the QFIM is proportional to the Fubini-Study metric:  
\begin{equation}
F_{ij}(\boldsymbol{\theta}) = 4g_{ij}(\boldsymbol{\theta}),
\end{equation}
where \( F(\boldsymbol{\theta}) \) is the QFIM and \( g(\boldsymbol{\theta}) \) represents the Fubini-Study metric.

Calculating \( F \) for arbitrary quantum states can be complex. For rank-\( r \) density matrices, defined as \(\rho(\boldsymbol{\theta}) = \sum_{n=1}^r \lambda_n |\lambda_n\rangle \langle \lambda_n|\), the entries of the Quantum Fisher Information Matrix (QFIM) are given by \cite{liu2020quantum}:
\begin{align}
F_{ab} &= \sum_{\lambda_i \in S} \frac{(\partial_a \lambda_i)(\partial_b \lambda_i)}{\lambda_i} 
+ \sum_{\lambda_i \in S} 4\lambda_i \, \text{Re} \left( \langle \partial_a \lambda_i | \partial_b \lambda_i \rangle \right) \nonumber \\ 
&\quad - \sum_{\lambda_i, \lambda_j \in S} \frac{8 \lambda_i \lambda_j}{\lambda_i + \lambda_j} \, \text{Re} \left( \langle \partial_a \lambda_i | \lambda_j \rangle \langle \lambda_j | \partial_b \lambda_i \rangle \right),
\end{align}
where \( S \) denotes the set of nonzero eigenvalues of \( \rho \).

For pure states, this expression simplifies to the Fubini-Study metric tensor, up to a constant factor:

\begin{equation}
F_{ab} = 4 \, \text{Re}\left[\langle \partial_a \psi | \partial_b \psi \rangle - \langle \partial_a \psi | \psi \rangle \langle \psi | \partial_b \psi \rangle \right]
\end{equation}

Particularizing for QAOA, this can be rewritten in the following form:

\begin{align}
F_{ab} = & 4\, \text{Re}\Bigg[\langle \psi_{a-1} | H_a \Bigg( \prod_{k=a}^{b-1} e^{i \theta_k H_k} \Bigg) H_b | \psi_{b-1} \rangle - \nonumber \\ 
& \quad - \langle \psi_{a-1} | H_a | \psi_{a-1} \rangle \langle \psi_{b-1} | H_b | \psi_{b-1} \rangle \Bigg].
\label{eq:F_ab_Pure}
\end{align}

This expression stems from the unitary parameter embedding defined by QAOA. In this framework, the Hamiltonian \( H_a =H_{zz} \) when \( a \) is odd and \( H_a = H_m \) when \( a \) is even, while \( \ket{\psi_{a-1}} \) represents the state obtained after evolving the initial state with \( a-1 \) parameters.

For diagonal elements, this simplifies further to:
\begin{equation}
F_{aa} = 4\, \text{Re}\left[ \langle \psi_{a-1} | H_a^2 | \psi_{a-1} \rangle - \langle \psi_{a-1} | H_a | \psi_{a-1} \rangle^2 \right].
\label{eq:F_ab_diagonal}
\end{equation}

To address mixed states arising from noisy circuits, we approximate the QFIM by the expression:
\begin{align}
F_{ab} &= 4\, \text{Re} \Big[\text{Tr}\left(\rho_{a-1} H_a \Bigg( \prod_{k=a}^{b-1} e^{i\theta_k H_k} \Bigg) H_b \rho_{b-1} \right) \nonumber \\
&\quad - \text{Tr}(\rho_{a-1} H_a) \cdot \text{Tr}(\rho_{b-1} H_b) \Big],
\label{eq:F_ab_approx}
\end{align}
and for diagonal elements:
\begin{equation}
F_{aa} = 4\, \text{Re} \left[ \text{Tr}(\rho_{a-1} H_a^2) - \left( \text{Tr}(\rho_{a-1} H_a) \right)^2 \right].
\label{eq:rfiag}
\end{equation}

Indeed, under the assumption of mild noise, mixed states can be expressed using the spectral decomposition as:
\begin{equation}
\rho = \lambda_0 |\psi\rangle \langle \psi| + \sum_{m=1}^{d} \lambda_m |\psi_m\rangle \langle \psi_m|,
\end{equation}
where the dominant eigenvector \( |\psi\rangle \) approximates the ideal quantum state \cite{koczor2021exponential, huggins2020virtual, koczor2022quantum}.

Thus, when the values of \( \lambda_m \) are small (low noise), the noisy state is close to the pure state, and Eqs.~\eqref{eq:F_ab_approx} and \eqref{eq:rfiag} provide a good approximation for the QFIM.

\subsection{Computation cost of QNG for QAOA}

The cost of computing the diagonal approximation of the Quantum Fisher Information Matrix (QFIM) for QAOA, as defined in Eq.~\eqref{eq:F_ab_diagonal}, depends on the number of expectation values required for each entry. Each expectation value generally requires a separate quantum circuit preparation. However, this cost can be reduced by grouping observables into sets of commuting operators. Observables within the same commuting group can be measured simultaneously using a single quantum circuit, thereby reducing the total number of distinct circuits required. This technique is a standard practice in the quantum computing community to reduce the resource overhead when estimating observables composed of many terms, such as Hamiltonians~\cite{Yen2020}.

In the worst-case scenario, for a Hamiltonian consisting of \( m \) non-commuting terms—where neither the terms of the Hamiltonian nor those of its square commute with each other—one must evaluate \( \mathcal{O}(m^2) \) expectation values to compute a single diagonal entry of the QFIM. This requires preparing and running \( \mathcal{O}(m^2) \) different quantum circuits. To estimate each expectation value with a precision of \( \mathcal{O}(\epsilon) \), \( \mathcal{O}(1/\epsilon^2) \) measurements are needed. Consequently, the total measurement cost for estimating one diagonal QFIM entry is \( \mathcal{O}(m^2/\epsilon^2) \).

If the Hamiltonians involved in QAOA have different numbers of terms, the cost of estimating the full diagonal QFIM becomes 
\( \mathcal{O}(P(m^2 + m'^2)/\epsilon^2) \), 
where \( m \) and \( m' \) are the numbers of non-commuting terms in the cost and mixing Hamiltonians, respectively, and \(P \) is the number of QAOA layers. However, as previously noted, this cost can be significantly reduced by grouping commuting observables and measuring them using the same circuit. This strategy is also employed in~\cite{stokes2020quantum} for estimating the block-diagonal approximation of the QFIM.

In our specific case of the transverse field Ising model (TFIM), the cost of computing the diagonal entries of the QFIM is substantially reduced due to the structure of the Hamiltonian. All terms in \( H_{zz} \) commute with each other, and the same holds for all terms in \( H_x \). This allows us to compute all required expectation values for each diagonal QFIM entry using only one quantum state preparation per group. As a result, only \( 2 P \) distinct quantum circuit preparations are needed for the diagonal approximation of the QFIM.

In contrast, computing the gradient requires \( 4m P \) different quantum circuits, where \( m \) is the number of terms in both the cost and mixing Hamiltonians.

On the other hand, computing the full QFIM is substantially more costly because it requires estimating off-diagonal terms between different layers. In general, estimating these terms necessitates the use of the Hadamard test, which requires an auxiliary qubit for each term involved in Eq.~\eqref{eq:F_ab_Pure}. To compute a single off-diagonal QFIM entry, one must perform \( \mathcal{O}(mm') \) different Hadamard tests, each requiring a separate quantum circuit with an auxiliary qubit, where \( m \) and \( m' \) are the number of terms in the respective Hamiltonians \( H_a \) and \( H_b \) from Eq.~\eqref{eq:F_ab_Pure}. Each such estimate with precision \( \mathcal{O}(\epsilon) \) requires \( \mathcal{O}(1/\epsilon^2) \) measurements, leading to a total cost of \( \mathcal{O}(mm'/\epsilon^2) \) measurements per off-diagonal entry.

Because the QFIM is symmetric, and assuming \( 2 P  \) parameters in the QAOA circuit, there are only \( P (2 P + 1) \) independent entries. Thus, the total cost of estimating the full QFIM in terms of measurements is upper-bounded by
\( \mathcal{O}( P^2 m^2/\epsilon^2) \), 
assuming \( m > m' \). If \( m' > m \), the roles can be swapped accordingly.

Importantly, the QFIM must be estimated at each step of the optimization process. Due to this unfavorable scaling, computing the full QFIM becomes impractical for large problem instances. By contrast, the diagonal approximation provides a more feasible alternative, offering potential improvements at a measurement cost similar to that of gradient estimation.

Nonetheless, gradient-based optimization remains out of reach for current near-term quantum devices. There is ongoing research aimed at reducing the complexity of estimating both gradients~\cite{Wierichs_2022, bowles2024backpropagationscalingparameterisedquantum} and the QFIM~\cite{gómezlurbe2025efficientprotocolestimatequantum}, which may enable these optimization strategies to become viable in practice in the near future.

Additionally, there has been rigorous analysis of the measurement overhead associated with variational quantum algorithms that utilize the QNG for optimization. In~\cite{PRXQuantum.2.030324}, the authors derive upper bounds on the number of measurements required to perform QNG at each optimization step for a given target precision. Notably, they show that, asymptotically, the cost of computing the QFIM becomes negligible compared to that of estimating the gradient, particularly as the number of iterations and qubits increases. In this asymptotic regime, QNG introduces only marginal additional overhead relative to standard Gradient Descent, while simultaneously improving convergence speed and overall performance. Furthermore, the authors propose a general strategy for optimally allocating measurement resources between the QFIM and the gradient to minimize total overhead.

\section{Rydberg atoms implementation}\label{Rydberg}

A quantum computer based on Rydberg atoms represents an ideal platform for testing the efficiency of our algorithm \cite{browaeys2020many, henriet2020quantum, saffman2016quantum}. Analog quantum computing leverages the natural evolution of a quantum system to solve specific problems. Instead of programming discrete quantum gates (digital quantum computing), physical parameters, such as magnetic fields or laser intensities, are directly manipulated to induce a continuous-time dynamic evolution. 

In the our case, we work with a \emph{register} of Rydberg atoms, which serve as the qubits. Analog quantum computing is realized by evolving this atomic register according to a specific Hamiltonian, while continuously adjusting parameters like detuning or laser intensity. Indeed, by acting on all atoms with the same global pulse, one can evolve the entire system with the following Hamiltonian:
\begin{equation}
H(t) = \sum_i \left( \frac{\Omega(t)}{2} \sigma^x_i - \delta(t) n_i + \sum_{j<i} U_{ij}n_i n_j \right)
\end{equation}
where $n_i=\frac{1}{2}(1+\sigma^z_i)$. Here, $\Omega(t)$ is the Rabi frequency (which determines the strength of the interaction and it is proportional to the amplitude of the laser field), $\delta(t)$ (which denotes the discrepancy between the qubit resonance and the field frequencies) and $U_{ij}$ denotes the blockade interaction parameter. This latter depends on the distance between qubits \(i\) and \(j\): $U_{ij}=C_6 / R_{ij}^6 $, where $C_6$ is a constant. By the continuously manipulating $\Omega(t)$ and $\delta(t)$, one can achieve a significant level of control over the system’s dynamics.

\begin{figure*}
   \centering
   \includegraphics[scale=0.3]{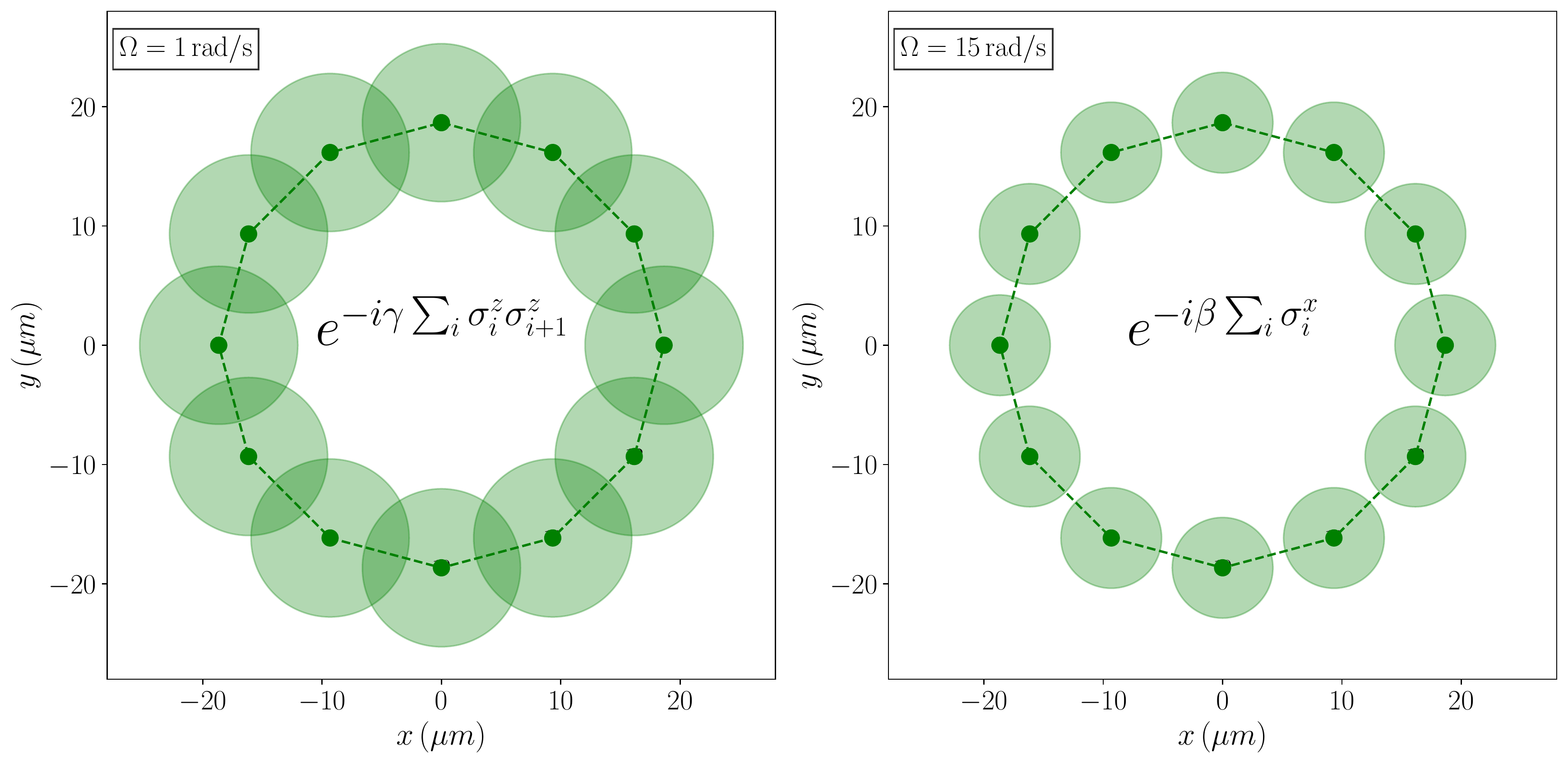}
   \caption{Effective implementation of the two time-evolution operators of the QAOA ansatz in a Rydberg atom register. In this example with $N=12$, we demonstrate how these unitaries can be realized by transitioning in and out of the blockade regime, adjusting the value of $\Omega$ from 1 rad/sec to 15 rad/sec.}
    \label{fig:Ryd_implem}
\end{figure*}

The evolution operators of the two Hamiltonians $H_{zz}$ and $H_m$ can be easily implemented by fixing the qubit register (thus fixing the distance between individual atoms $R_{ij}$) and varying $\Omega(t)$ to enter and exit the blockade regime. Inside the blockade regime only the interaction terms $n_i n_{i+1}$ will be relevant, whereas outside this regime $\sigma_x$ will be the dominant factor. The blockade regime is determined by the Rydberg radius:
\begin{equation}
R_b = \left( \frac{C_6}{\Omega(t)} \right)^{1/6}.
\end{equation}
In both regimes, we set the detuning value \(\delta(t) = U_{i,i+1}\).
In fact, expanding the interaction term \( U_{i,i+1} n_i n_{i+1} \) reveals that it contains both \( \sigma^z_i \sigma^z_{i+1} \) and \( \sigma^z_i \) terms. The latter can be canceled by this choice of detuning.
An example of this implementation for $N=12$ can be seen in Fig.~\ref{fig:Ryd_implem}. \\
This work focuses on the simulation of pulse sequences (including noise and imperfections) using the Pulser library~\cite{henriet_pulser_docs_2020,silverio2022pulser}, a Python-based library specifically designed to simulate the dynamics of Rydberg atom systems by numerically solving the full time-dependent Schrödinger equation in the Hilbert space of the system (i.e. full state evolution).
These factors are intrinsic to real-world quantum devices and must be accurately replicated in simulations to devise effective strategies for addressing them. A comprehensive explanation of these phenomena in Rydberg platforms is provided in \cite{de2018analysis}. Here, we will briefly outline the aspects considered in this study.
\begin{enumerate}
    \item \textbf{SPAM (State Preparation And Measurement) Errors.}
    \begin{itemize}
        \item \textbf{State Preparation Errors.} Initial attempts to prepare the state may occasionally fail to capture all atoms in the ground state. This is modeled using a probability, $p$, denoting the likelihood of an atom not being in the intended state during preparation.
        \item \textbf{Measurement Errors.} Detection inaccuracies contribute to errors. These include falsely identifying an atom in the ground state as excited (``false positives'') and vice versa (``false negatives'').
    \end{itemize}
    
    \item \textbf{Laser Noises.}
     Laser properties such as frequency and amplitude fluctuations impact their efficacy in addressing atomic level transitions.
    \begin{itemize}
       
        \item \textbf{Doppler Effect.} Thermal motion of atoms causes Doppler shifts in the laser frequencies, altering the detuning frequency.
        \item \textbf{Laser Waist.} The Gaussian profile of laser amplitude means atoms at the edges of the beam experience slightly lower amplitudes compared to those at the focal point.
        \item \textbf{Amplitude Fluctuations.} Laser amplitude varies from pulse to pulse, influencing the consistency of laser operations.
    \end{itemize}
\end{enumerate}

\section{Numerical results}

In this section, we assessed the performance of the QAOA with QNG and Vanilla gradient descent by analyzing their effectiveness in preparing the ground state of the TFIM by using Rydberg atomos platform. Our numerical studies included comparisons of convergence rates in reaching the true ground state from multiple random parameter initializations, both with and without noise, as well as the number of steps required to achieve convergence.

\subsection{Ground state preparation}

As explained in the previous sections, the QAOA algorithm adheres to a specific protocol, which in our case involves initializing the state in $\ket{+}^{\otimes N}$ and evolving it using the Hamiltonians $H_{zz}$ and $H_m$. By strictly following this procedure, we find numerical evidence that the ground state is attained at a circuit depth \( P = \lfloor N/2 \rfloor \), as verified in Sec.~\ref{Application to TFIM}.

However, as previously mentioned, it is not feasible to evolve precisely using those two Hamiltonians since the interaction term, even though it should be negligible outside the Blockade regime, remains present and might affect the accuracy and the number of layers required to reach the ground state. The same applies within the blockade regime, where $\Omega(t)$, although small, remains present. 

\begin{figure*}
\centering
\hspace{0.65cm}
    \small Noiseless \hspace{3.6cm} \small   Laser noise \hspace{3.4cm} \small SPAM errors \\
    \centering
    \includegraphics[scale=0.35]{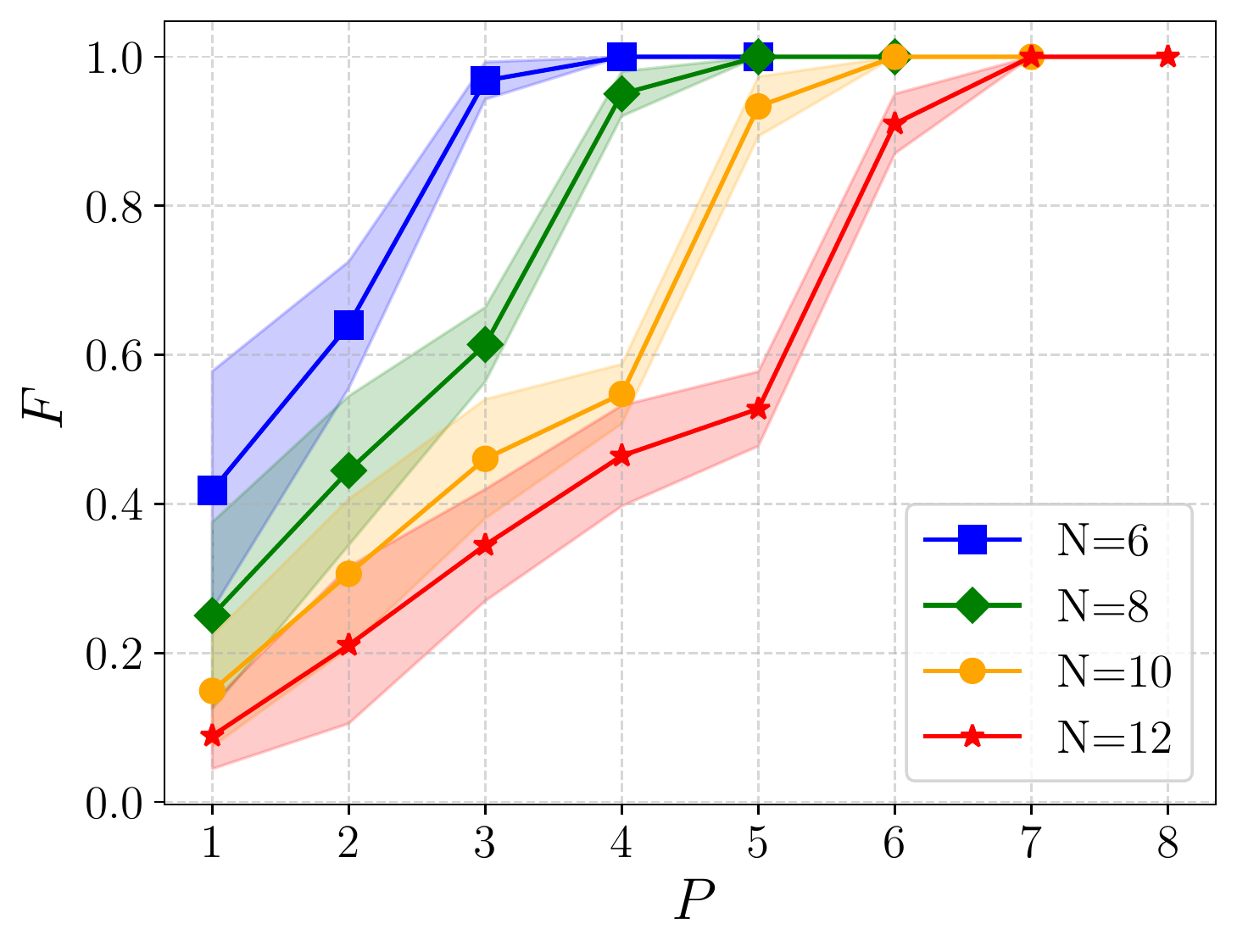}
    \includegraphics[scale=0.35]{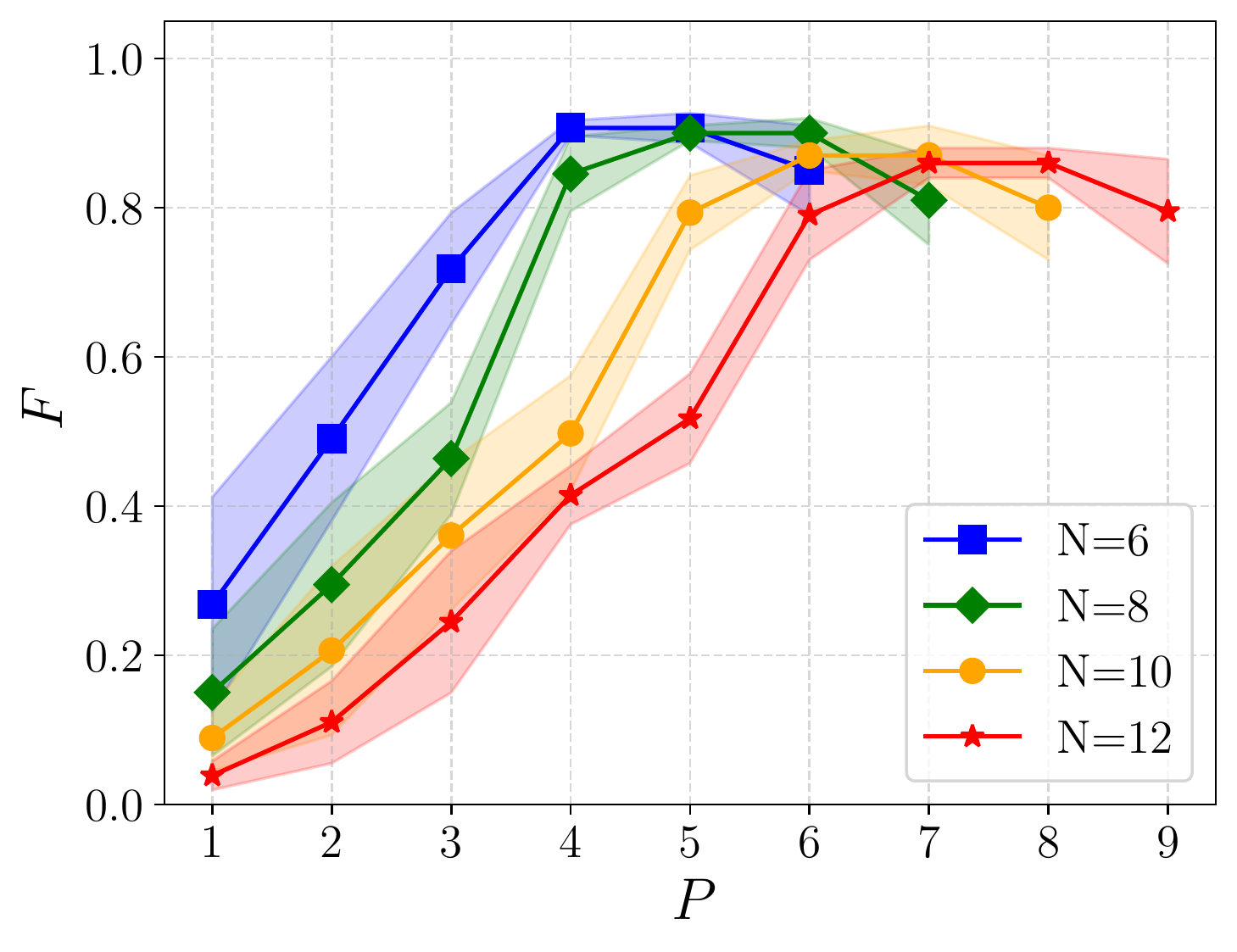}
    \includegraphics[scale=0.35]{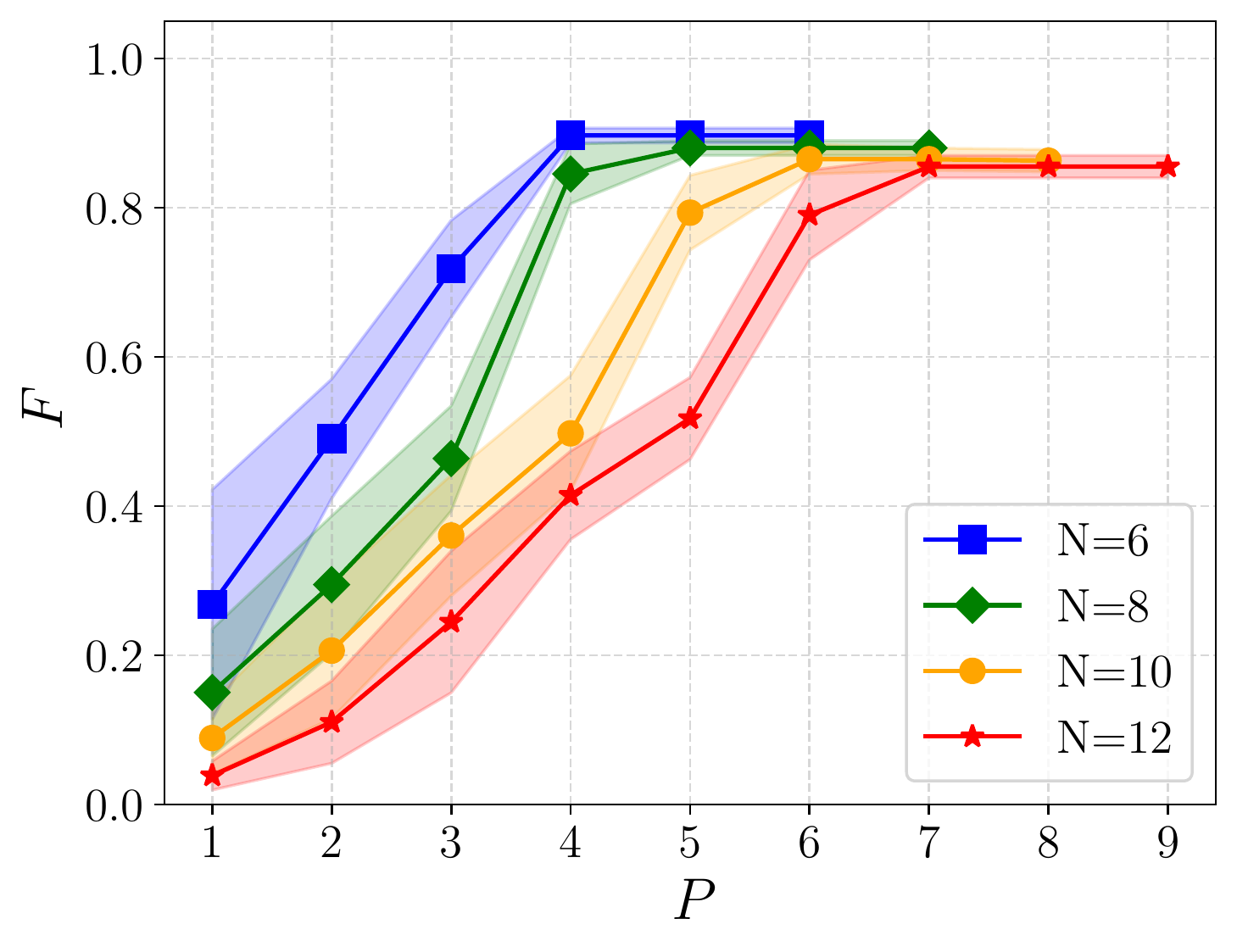}
    \caption{Growth of fidelity as a function of depth \(P\) for various system sizes (\(N = [6, 8, 10, 12]\)) and under three different scenarios: noiseless, laser noise, and SPAM errors. We used the Quantum Natural Gradient as the optimizer. It is noteworthy that, in each case, the highest value of \(F\) is achieved when \(P = \lfloor N/2 \rfloor + 1\).}
    \label{fig:Fidelities_vs_P}
  
\end{figure*}

We aimed at exploring this aspect by plotting how fidelity varies with the number of layers \(P\) for different system sizes, both in the presence and absence of noise (see Fig.~\ref{fig:Fidelities_vs_P}). We opted to present the fidelity trends because, in some cases, final states with nearly identical energies may still exhibit low fidelity values. As our primary goal is state preparation, our focus is on how closely the achieved state overlaps with the true ground state. The fidelity between a mixed state and a pure state is defined as \(F = \bra{\psi_g}\rho\ket{\psi_g}\), where \(\psi_g\) is the ground state vector determined through exact diagonalization.
According to the simulation results, it is clear that in the noiseless scenario, the ground state is achieved with an accuracy  \(\delta E_{opt}<10^{-9}\) when \(P = \lfloor N/2 \rfloor + 1\), where \( \delta E_{\text{opt}} \) represents the best accuracy obtained after optimization. This accuracy is calculated as the difference between the energy value obtained by the algorithm at convergence and the true value (see Eq.~\eqref{eq:accuaracy}). 

When two different sources of noise are introduced, we find that the optimal fidelity is consistently reached at \(P = \lfloor N/2 \rfloor + 1\), with accuracy ranging from $10^{-7}$ to $10^{-4}$, as \(N\) changes. Additionally, the plots include fidelity values for two extra layers to evaluate whether this might help the algorithm in achieving the desired ground state. However, the results indicate that adding further layers does not provide any advantage: in the case of laser noise, the fidelity value stabilizes at the next layer after \(P = \lfloor N/2 \rfloor + 1\) (though with increased variance), followed by a subsequent decrease. For spam errors, both the fidelity and its variance remain constant for \(P \geq \lfloor N/2 \rfloor + 1\). This can be easily understood as adding parameters does not increase noise, unlike the situation with laser noise evolution.

\subsection{Comparison of QNG against Vanilla gradient descent}

We evaluated the performance of QNG and standard gradient descent in finding the true ground state of 1D TFIM by testing both methods across various system sizes, setting $h=0.5$ and $J=1$. 
The simulations were conducted under three different conditions: noiseless, with laser noise, and with SPAM errors.
The maximum number of iterations allowed for the optimization was set to $5000$ while the convergence criterion to stop the optimization procedure, according to equation \ref{conv}, was \( \epsilon_{stop} = 10^{-12} \). The learning rate $\eta$ was selected iteratively by conducting some trials with a fixed $N$, and choosing the best-performing values for both methods.

Furthermore, the QNG was computed using both the full Quantum Fisher Information Matrix (QFIM) and the diagonal approximation (which utilizes only the diagonal elements of the QFIM). This comparison is crucial for understanding the scenarios in which it is feasible to approximate the QFIM, thus calculating fewer elements. This is significant because the QFIM must be inverted during the parameter update process, which incurs in additional computational overhead.
\subsubsection{Noiseless Case}
\begin{figure}[]
    \centering
    \textbf{Noiseless} \\ \vspace{0.2cm}
        \includegraphics[scale=0.28]{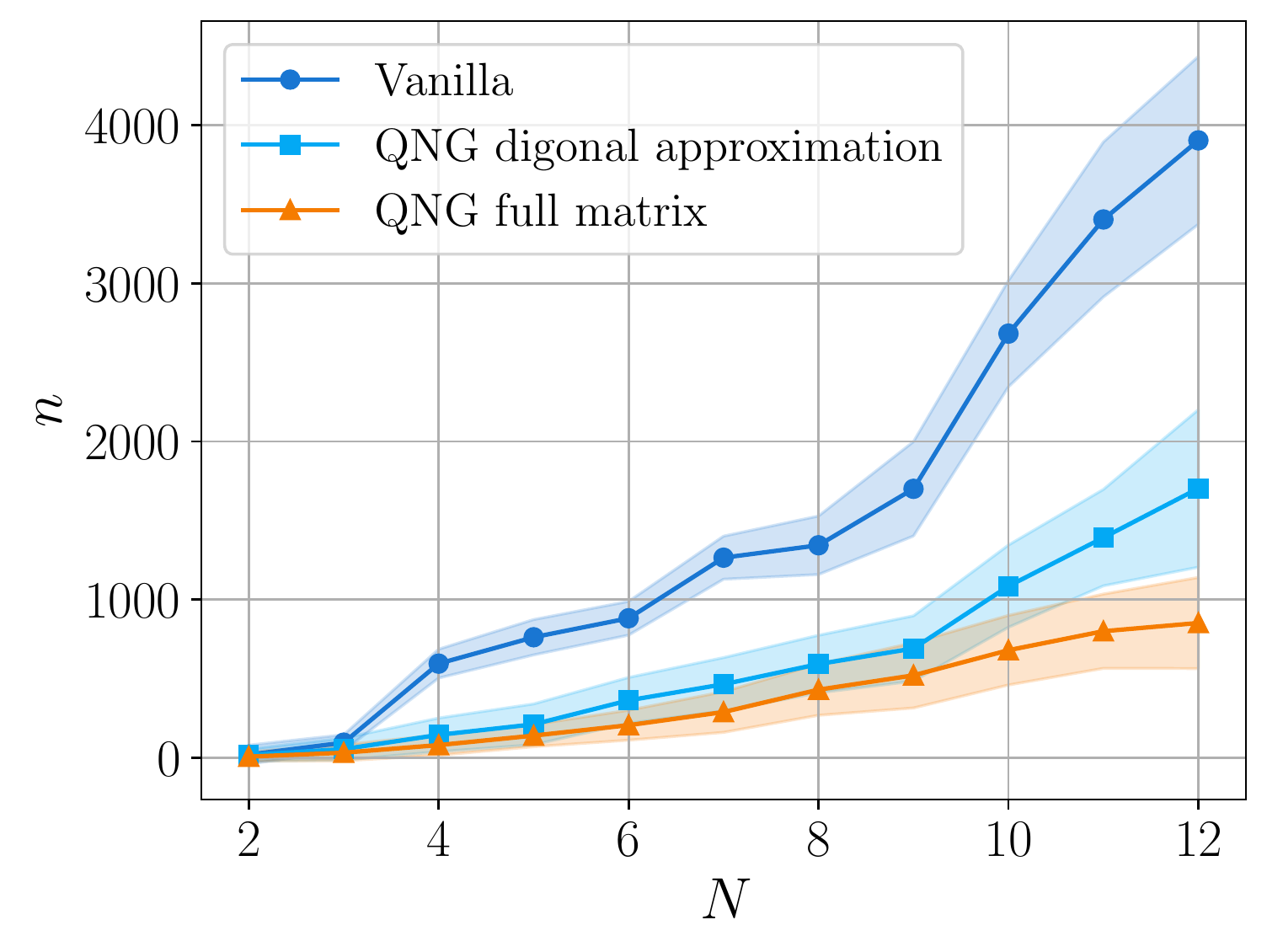}
    \centering
        \includegraphics[scale=0.28]{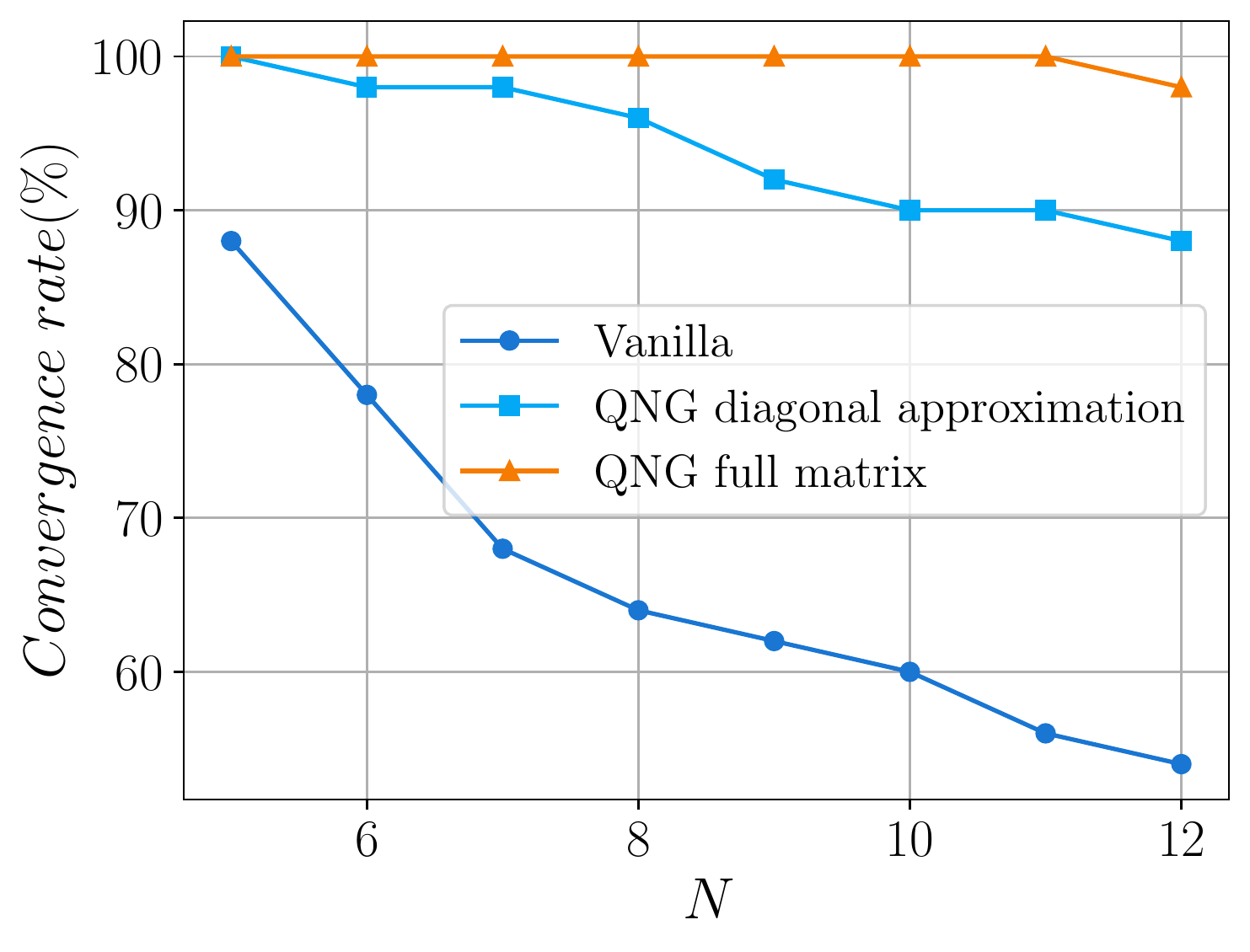}
  \caption{\small Comparison of performance metrics for the Vanilla optimizer and Quantum Natural Gradient (QNG) methods (both diagonal approximation and full matrix) in the noiseless case. The left panel shows the average number of steps, \( n \), required to reach the ground state as a function of the system size \( N = [2, 12] \). The right panel displays the convergence rate of each method, reflecting the percentage of successful completions in reaching the ground state. Results demonstrate the growing performance gap between Vanilla and QNG as \( N \) increases, with the QNG methods maintaining higher success rates and lower step requirements, particularly the QNG full matrix approach.}

 \label{fig:noiseless}
\end{figure}

The noiseless case results are shown in Fig.~\ref{fig:noiseless}, where two primary figures of merit were analyzed: the \emph{average number of steps} required to reach the ground state and the \emph{convergence rate}, representing the percentage of successful algorithm completions in finding the ground state. Success in reaching the ground state is defined by achieving an accuracy  \(\delta E_{opt} < 10^{-9}\), where \( \delta E_{\text{opt}} \) represents the best accuracy obtained after optimization.
These values were calculated by running the algorithm from 50 randomly chosen initial points. 

An analysis of the average number of steps reveals a performance gap between Vanilla and QNG methods (both diagonal approximation and full matrix). Initially, this gap is small for smaller system sizes but grows as \(N\) increases. Notably, within the simulated sizes, the difference between QNG with the diagonal approximation and the full matrix appears minimal at first but becomes more pronounced starting from $N=9$.

Fig.~\ref{fig:noiseless} also indicates an increase in variance as the number of parameters \(P\) increases, attributable to the larger variety of initial points randomly selected. Here, the distinction between the classical optimizer and its quantum counterparts becomes apparent as well. Moreover, it is important to highlight that the performance of the Vanilla and QNG diagonal approximation is even worse than that of the full Matrix approach, as they are less robust against random initialization.

In terms of convergence rate (Fig.~\ref{fig:noiseless}, right panel), it varies significantly among the methods. Vanilla's success rate declines from \(88\%\) at \(N=5\) to \(56\%\) at \(N=12\). In contrast, QNG without approximation maintains nearly \(100\%\) success up to \(N=11\), with only a slight reduction to \(98\%\) at \(N=12\). The gap between the two quantum methods is more noticeable here, as the success rate for QNG with the diagonal approximation gradually decreases, reaching \(88\%\) at \(N=12\).

\begin{figure*}
      \centering
    \textbf{Noiseless accuracies} \vspace{0.2cm } \\
    \includegraphics[scale=0.3]{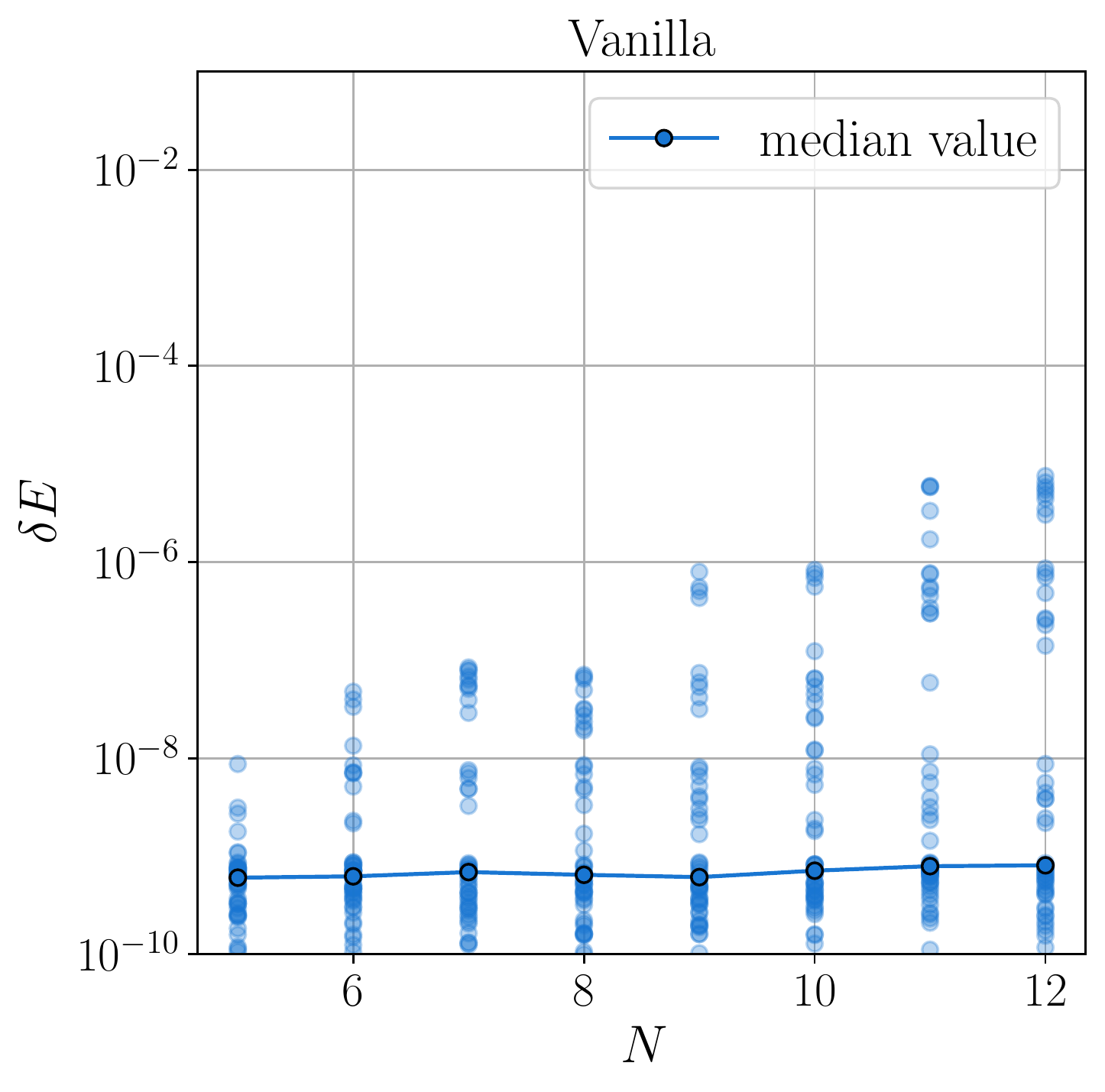}
     \includegraphics[scale=0.3]{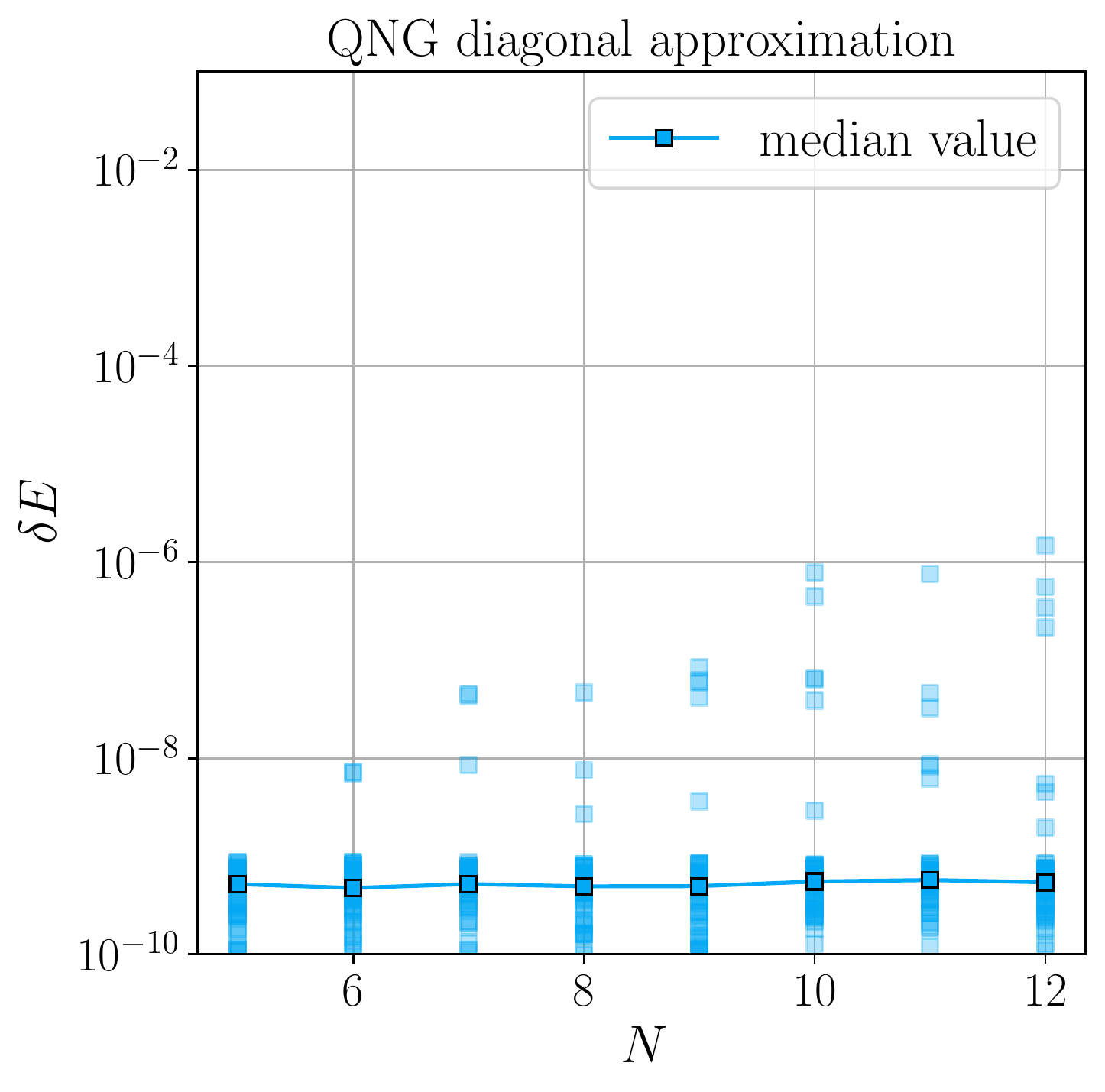}
      \includegraphics[scale=0.3]{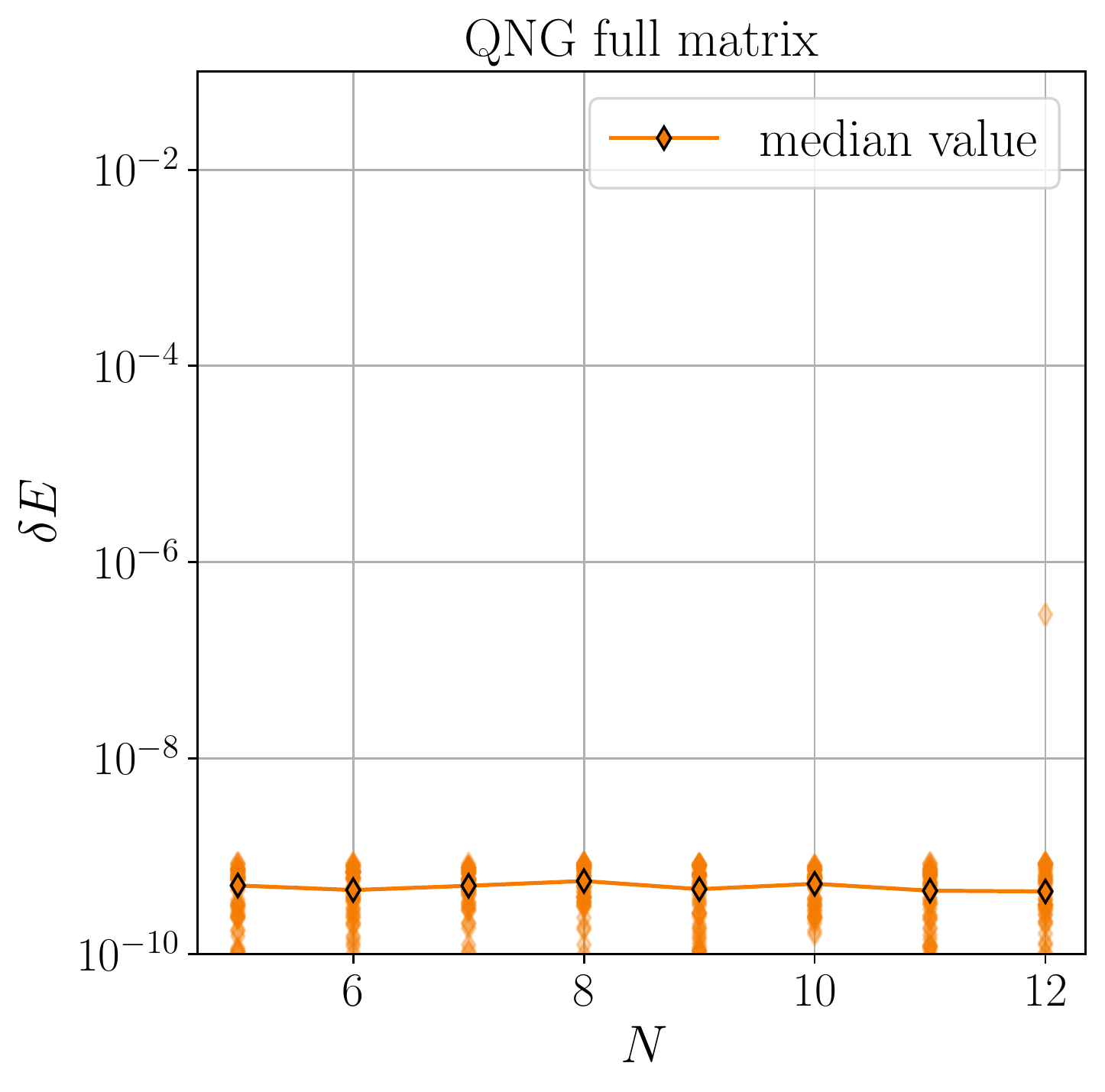}
    \caption{Comparison of noiseless accuracies for the Vanilla, QNG diagonal approximation, and QNG full matrix methods as a function of system size \( N \). Each panel displays the distribution of 50 accuracy values (measured as \( \delta E \)) for each $N$, obtained from different runs, with the median value of accuracies represented by solid lines. The most populated accuracy sector is at \( 10^{-10} \), indicating successful convergence to the ground state, especially for the QNG Full Matrix which is more consistent as $N$ increases.}
    \label{fig:acc_noiseless}
\end{figure*}

Fig.~\ref{fig:acc_noiseless} shows the distribution of 50 accuracy values as a function of system size \( N \). The most frequent accuracy is \(10^{-10}\), matching the percentage of successful outcomes shown in the right panel of Fig.~\ref{fig:noiseless}. The accuracy range for Vanilla widens as \(N\) increases, reaching approximately \(10^{-6}\) by \(N=12\), indicating a performance decline. By contrast, the QNG methods, especially the full matrix, demonstrate a significantly smaller spread, suggesting stability and robustness. This distribution highlights the higher variability and lower consistency of the Vanilla method, particularly at larger \(N\), while QNG methods remain narrowly distributed around lower values.
Finally, a comparison of the two QNG methods shows that while the full QNG consistently yields lower errors, the diagonal approximation offers a practical trade-off. Despite a slight decrease in accuracy as \(N\) increases, the diagonal approximation still outperforms Vanilla, making it a viable choice when computational resources are limited. 

\begin{figure}
        \centering
 \hspace{.7cm} \textbf{Laser noise} \hspace{1.9cm} \textbf{SPAM errors} \\ \vspace{0.2cm}
        \includegraphics[scale=0.275]{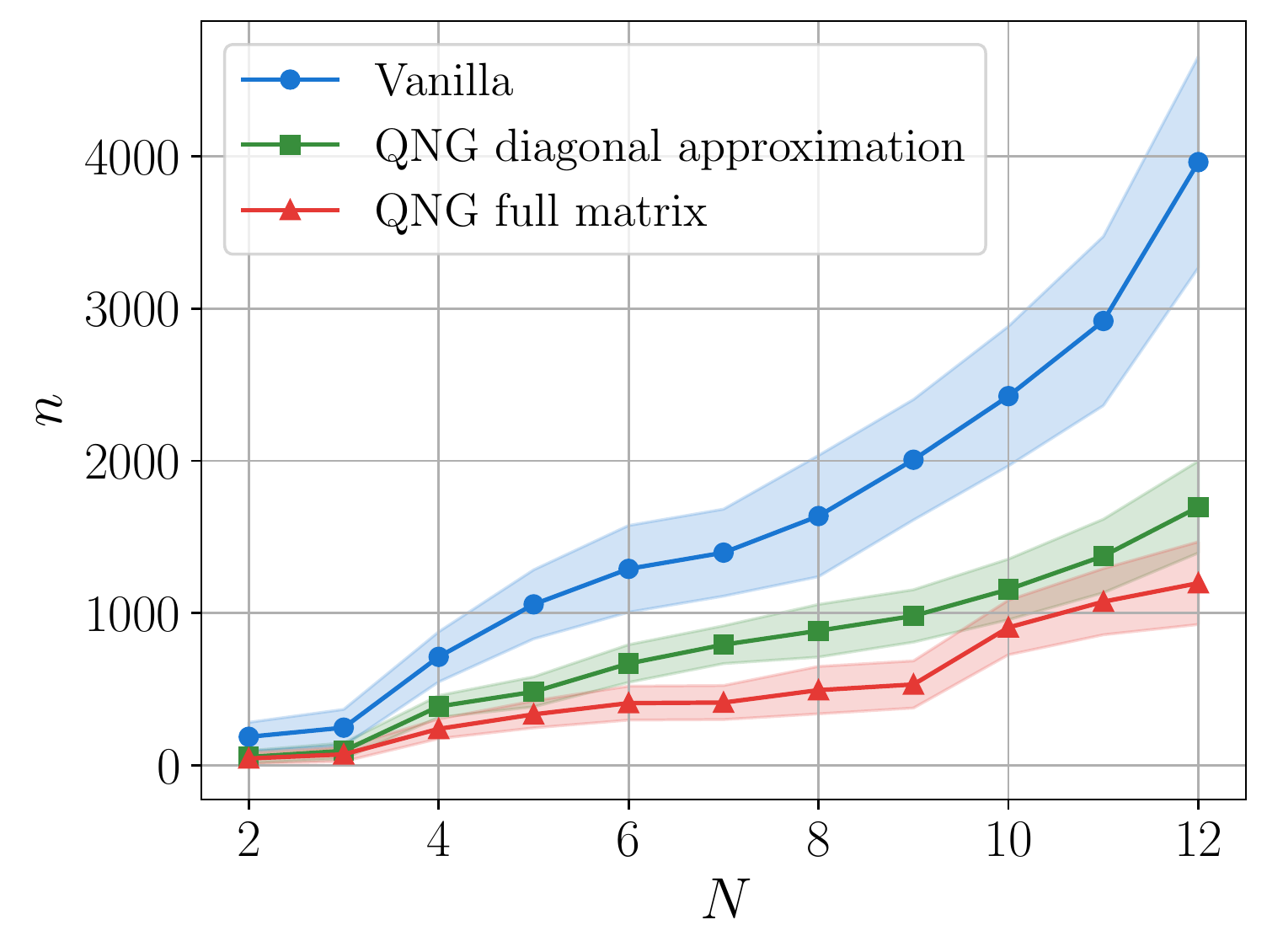}
       \includegraphics[scale=0.275]{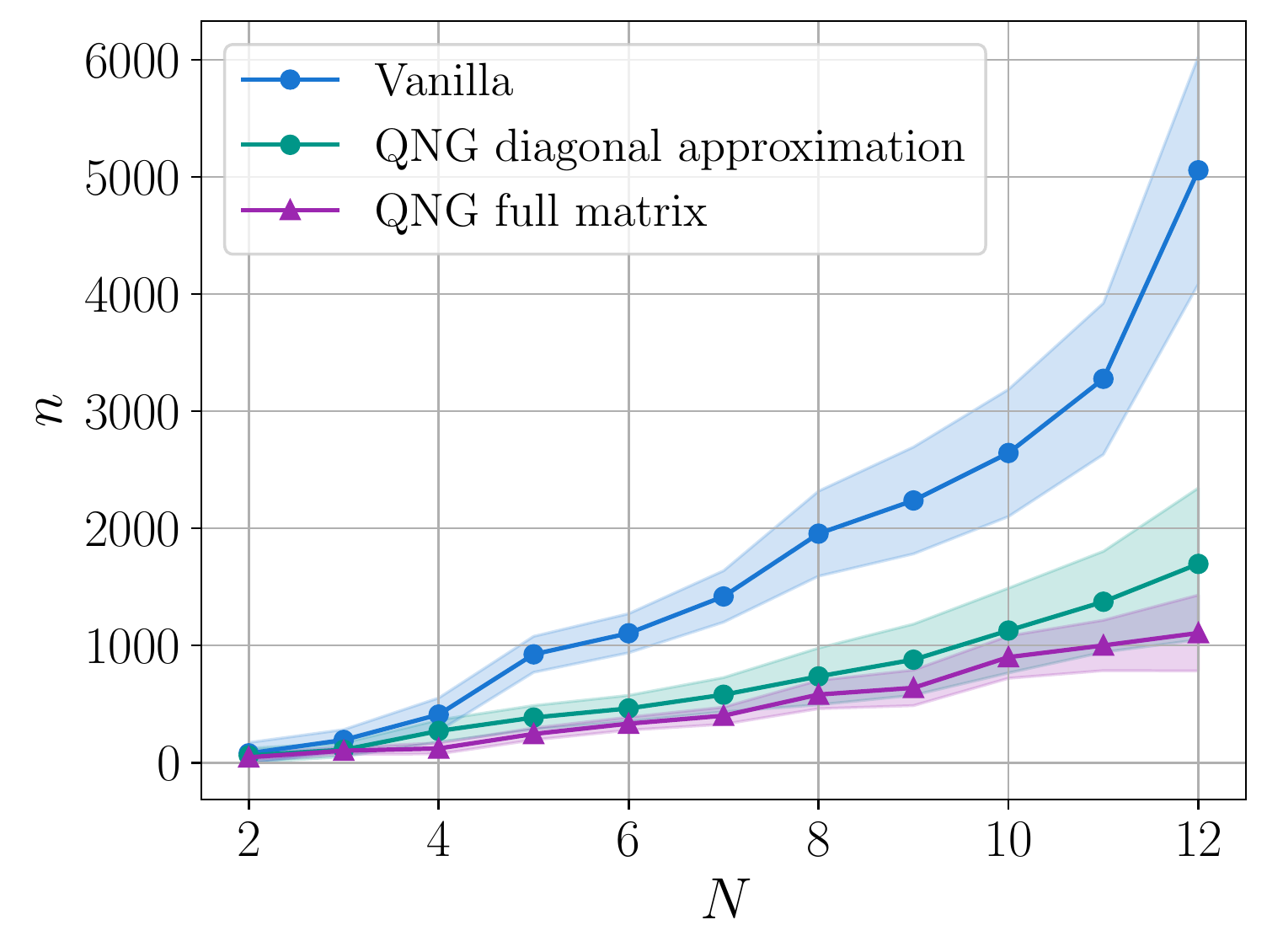}
\caption{The effect of laser noise (left) and SPAM errors (right) on the average number of steps \(n\) required to reach the cost function minimum as a function of system size \(N\). The three methods compared are Vanilla (blue), QNG with diagonal approximation (green), and QNG with full matrix (red/purple). The QNG full matrix method shows the best performance, requiring fewer steps and demonstrating lower sensitivity to noise, while the Vanilla method performs the worst, with greater sensitivity and rapid growth of \(n\) as \(N\) increases.}

\label{fig:n_laser_spam}
\end{figure}

\subsubsection{Laser noise and SPAM errors cases}
In the following subsection, we discuss the numerical results obtained when including both Laser noise and SPAM errors in the simulations, using the Python library \texttt{pulser} \cite{henriet_pulser_docs_2020}, which was developed to simulate Rydberg atom arrays and allows for precise control over parameters characterizing different sources of noise. Specifically, we kept the default temperature \texttt{T} \(= 50 \, \mu \text{K} \) to account for Doppler effects, as well as the Gaussian standard deviation for the laser waist, \texttt{laser\_waist} \(= 175 \, \mu \text{m}\), and amplitude fluctuations, \texttt{amp\_sigma} \(= 0.05\), which are representative values for current experimental setups (see also the \texttt{pulser} documentation \cite{henriet_pulser_docs_2020} for further details).
Also for the SPAM errors, we used the default values: \texttt{$\upeta=0.005$ }, \texttt{$\upepsilon = 0.01$}, and \texttt{$\upepsilon' = 0.05$}, which represent the probability of state preparation errors and the probabilities of measurement errors (false positives and false negatives) respectively.

\begin{figure*}
    \centering
    \textbf{Laser noise accuracies}  \vspace{0.2cm } \\
        \includegraphics[scale=0.3]{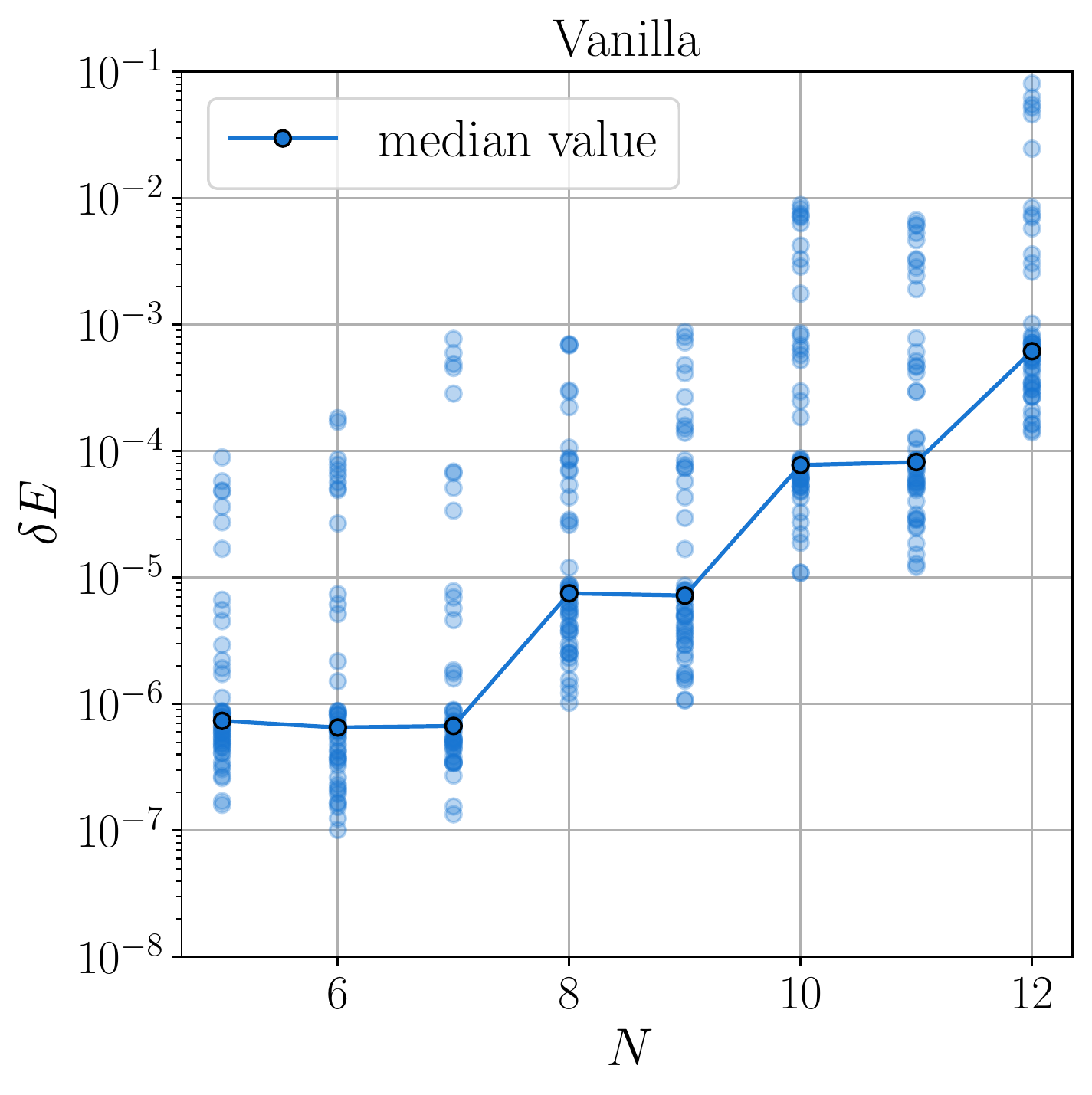} 
         \includegraphics[scale=0.3]{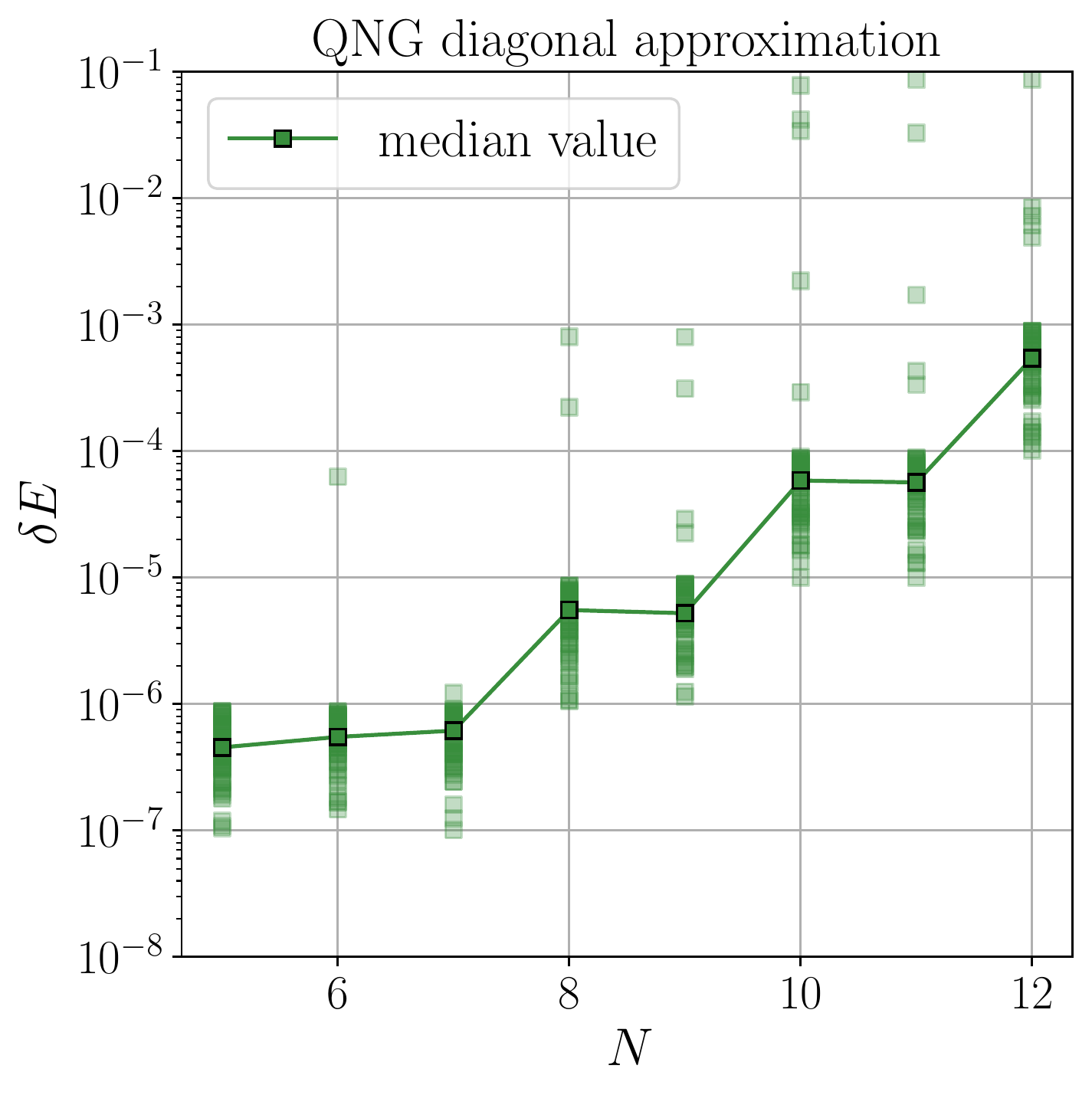} 
         \includegraphics[scale=0.3]{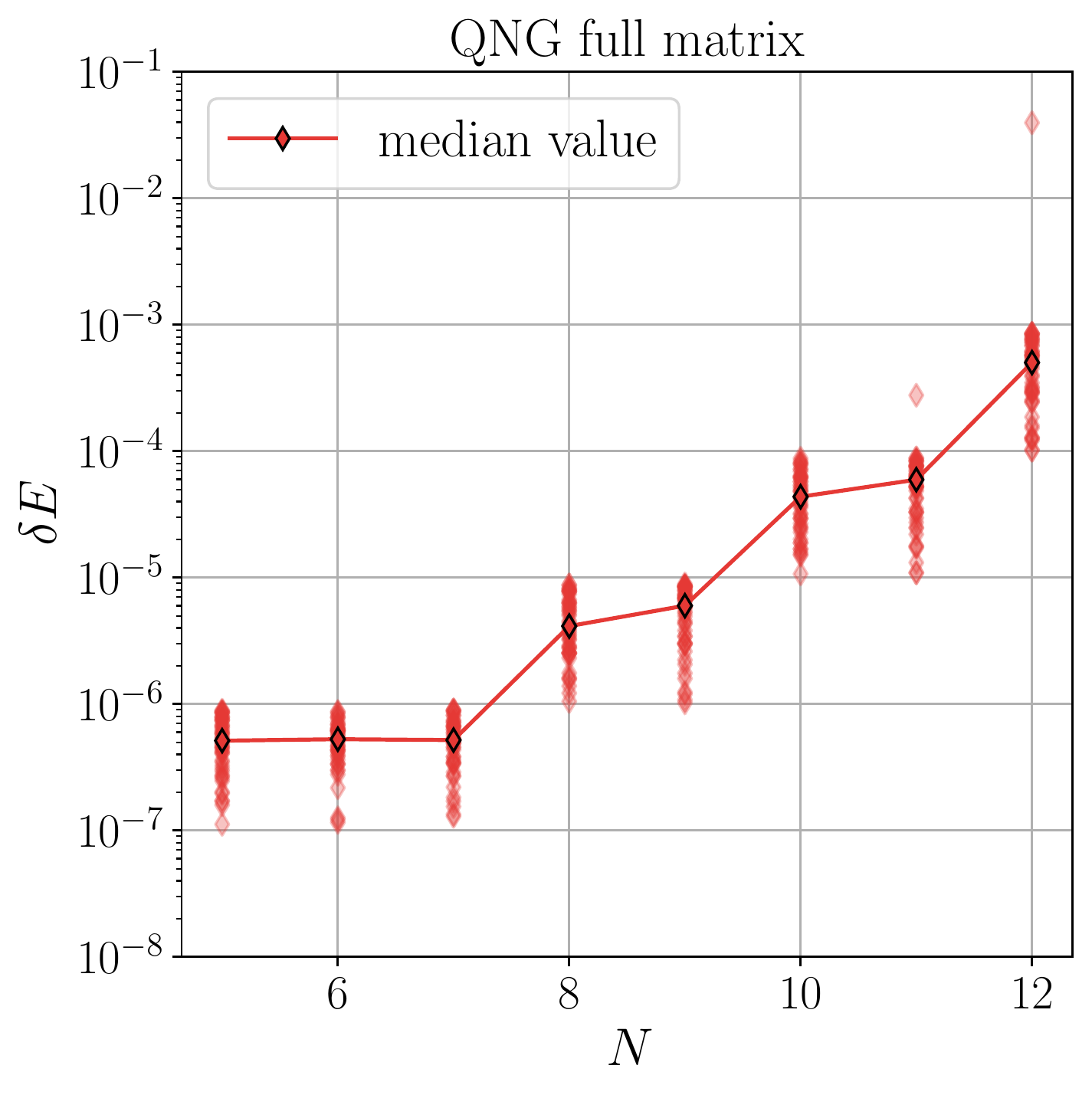} 
        \\
        \textbf{SPAM error accuracies} \vspace{0.2cm } \\
                \includegraphics[scale=0.3]{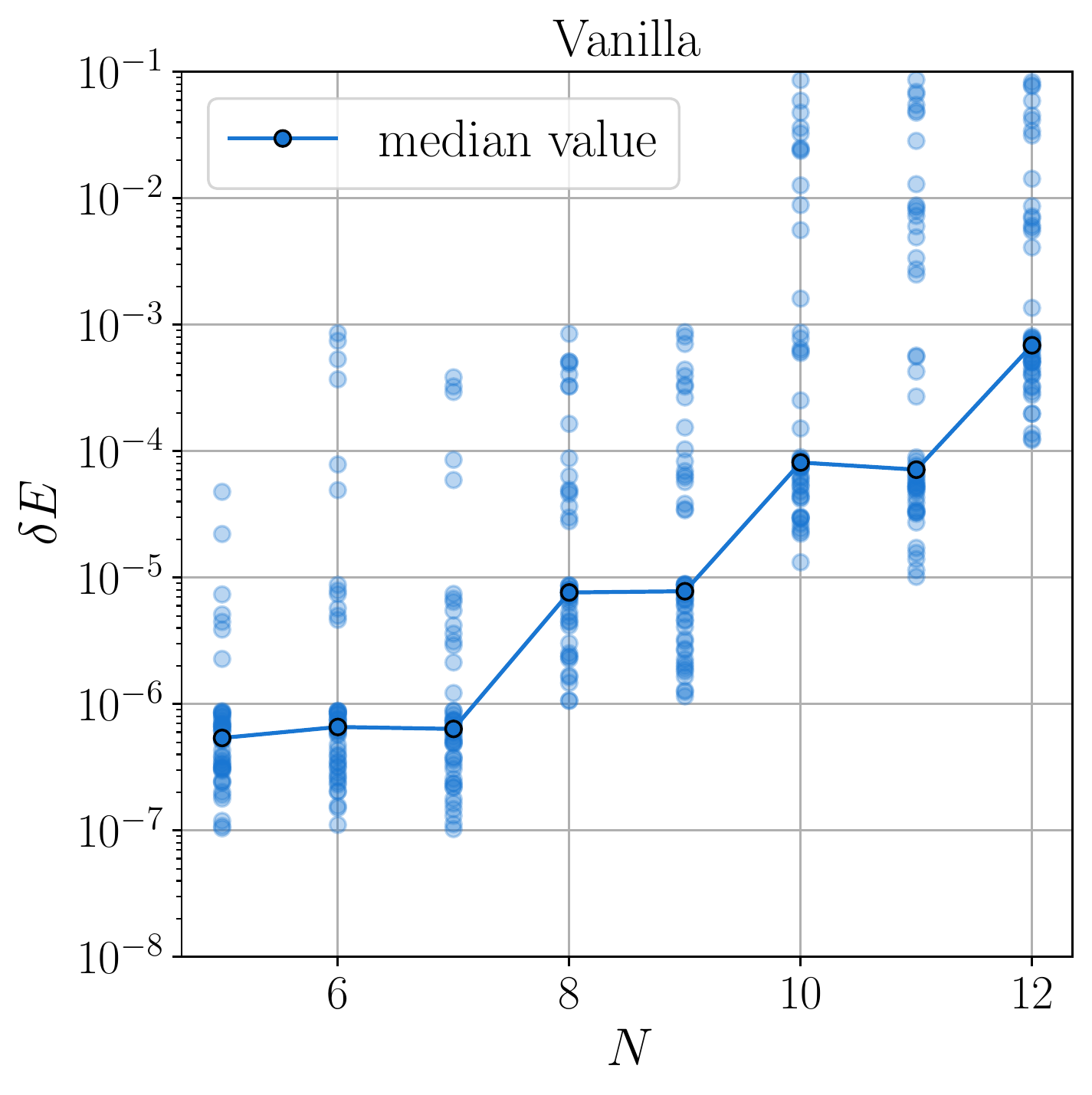}
                 \includegraphics[scale=0.3]{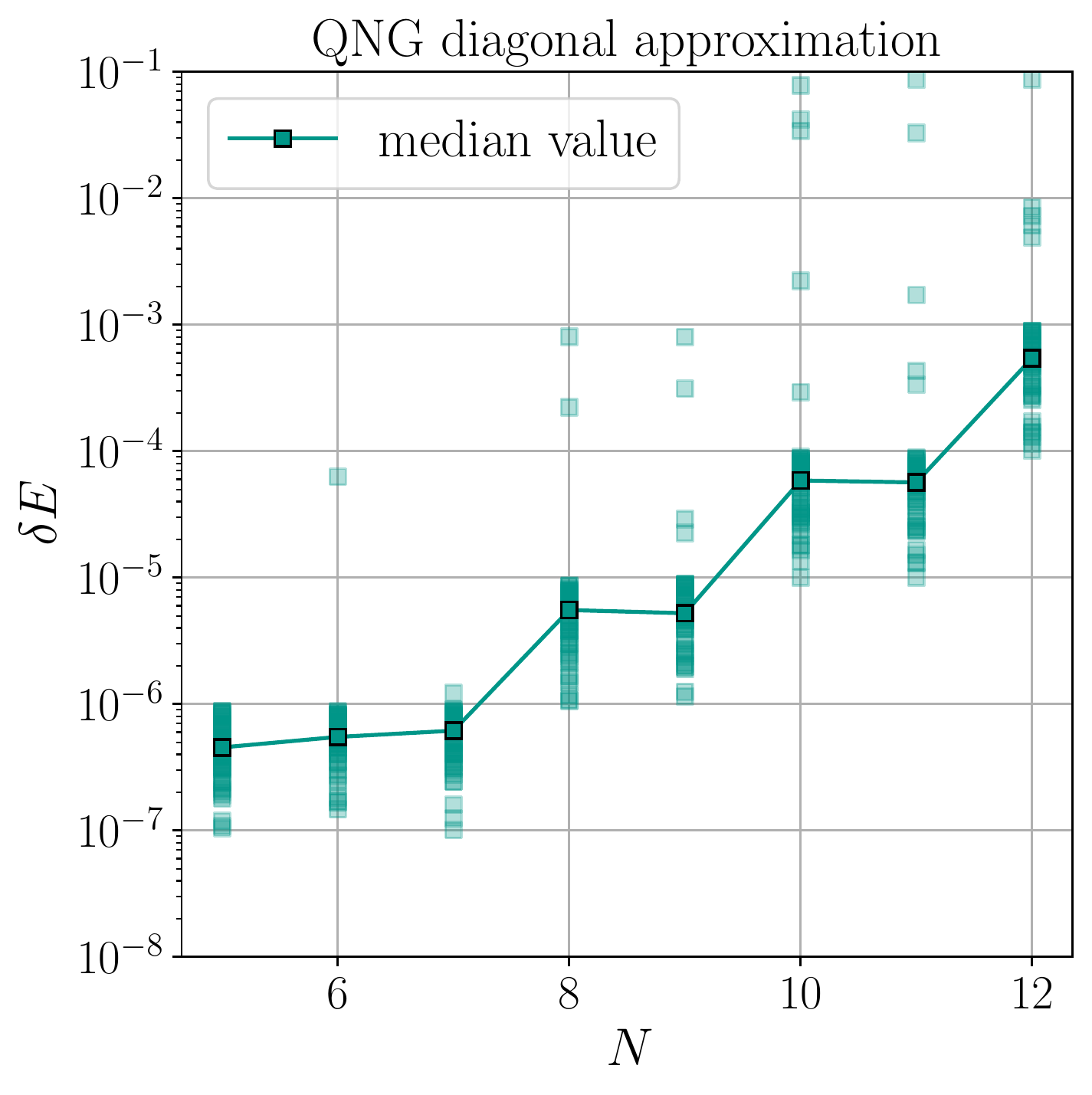}
                \includegraphics[scale=0.3]{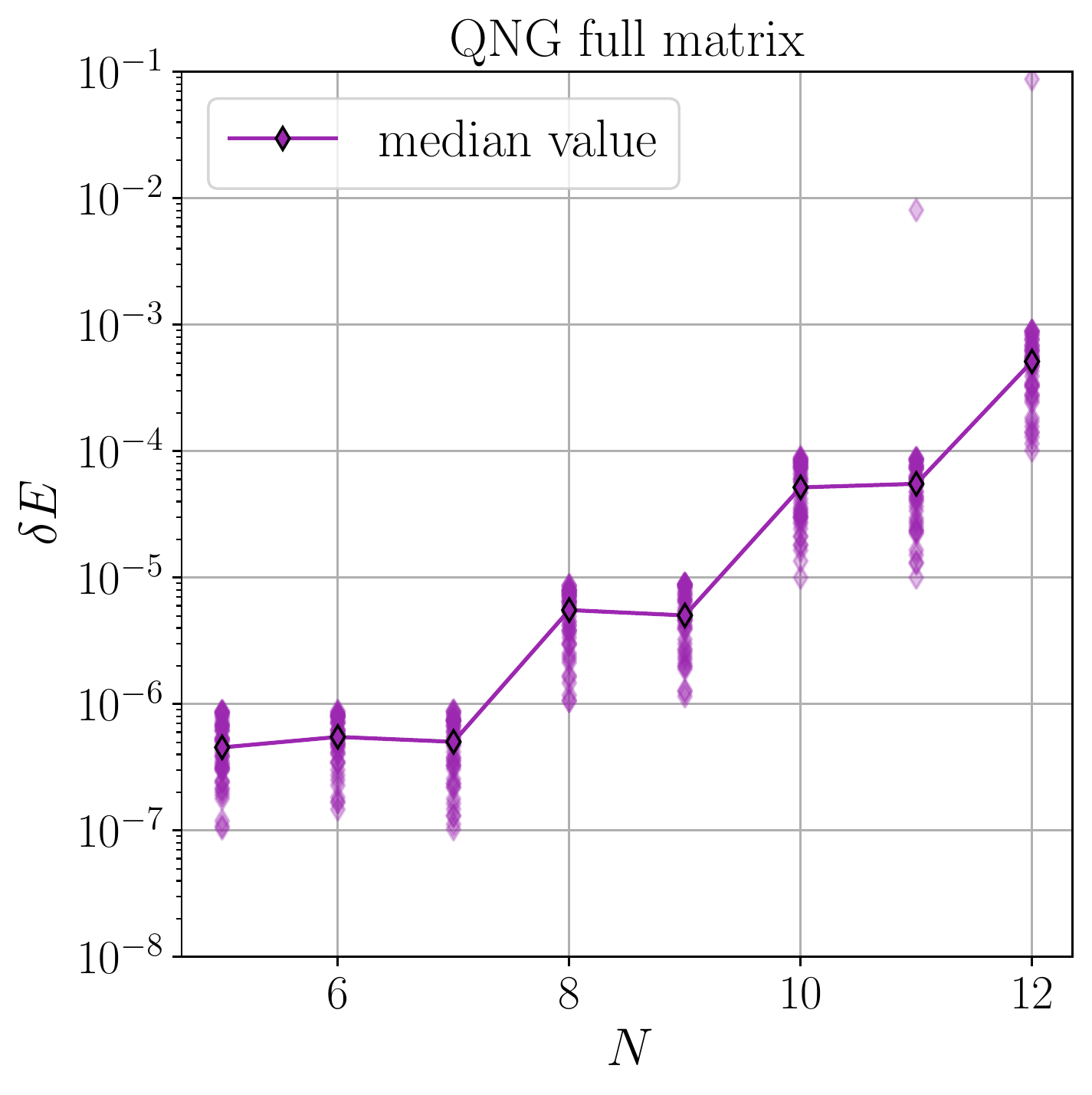}
\caption{The accuracy distribution \(\delta E\) across 50 runs as a function of system size \(N\) for three methods—Vanilla (left), QNG with diagonal approximation (center), and QNG with full matrix (right)—under laser noise (top row) and SPAM errors (bottom row). The QNG full matrix method achieves the lowest \(\delta E\) with minimal variance, indicating superior stability and robustness to noise, while the Vanilla method shows significant growth in \(\delta E\) and wide distributions, highlighting its high sensitivity to noise.}

    \label{fig:accuracies_noisy}
\end{figure*}

Figs.~\ref{fig:n_laser_spam} and \ref{fig:accuracies_noisy} illustrate the effects of Laser noise and SPAM errors on the performance of three methods—Vanilla, QNG with diagonal approximation, and QNG with full matrix—as a function of system size \(N\).

Fig.~\ref{fig:n_laser_spam} highlights the average number of steps \(n\) required to reach the convergence.
Under both noise types, the Vanilla method (blue curves) exhibits a steep increase in \(n\) as \(N\) grows, along with a wide confidence interval, indicating high sensitivity to noise and inconsistent performance. The QNG diagonal approximation (green curves) provides a notable improvement by reducing the number of steps and showing slower growth of \(n\) with \(N\). Its narrower confidence interval suggests moderate robustness to noise. The QNG full matrix method (red/purple curves) demonstrates the best performance, with \(n\) remaining relatively low and growing only slightly with \(N\). Additionally, its narrow confidence interval reflects strong resilience to both laser noise and SPAM errors. These results underline the efficiency and stability of the QNG methods, particularly the full matrix variant, in mitigating noise-related challenges during optimization

Fig.~\ref{fig:accuracies_noisy} provides further insights by presenting the accuracy distribution \(\delta E\) across 50 runs for the same three methods. For laser noise (top row), the Vanilla method (left) shows a sharp increase in \(\delta E\) with \(N\), along with a wide distribution spanning several orders of magnitude. This indicates its pronounced sensitivity to noise and poor reliability. In contrast, the QNG diagonal approximation (center) reduces variance and the median is slightly below compared to the Vanilla one, reflecting enhanced robustness to laser noise. The QNG full matrix method (right) delivers superior performance, maintaining low \(\delta E\) values with minimal spreadness of them, demonstrating very good stability and accuracy.

\section{Quantum Circuit Implementation and IBM Noise Model}\label{IBM} 

In this section, we describe the quantum circuit implementation for the QAOA and the integration of a realistic noise model based on calibration data from the IBM quantum device, \texttt{ibm\_sherbrooke}. This noise modeling approach enables a more realistic assessment of QAOA’s performance for the TFIM under hardware constraints. The use of device-specific noise models provides valuable insights into how noise affects the algorithm’s convergence and accuracy in practical scenarios.

\subsection{Circuit Construction and Hamiltonian Implementation}
For our QAOA implementation, we employed the Pennylane library \cite{bergholm2022pennylaneautomaticdifferentiationhybrid}, which enables efficient construction, simulation, and optimization of quantum circuits. The QAOA ansatz for solving the ground state of the Transverse Field Ising Model (TFIM) consists of alternating applications of two Hamiltonians, as described in Sec.~\ref{QAOACIRC}.

In this circuit-based implementation, we use Pennylane's Ising ZZ two-qubit gates:

\[
ZZ(\gamma) = \exp\left(-i \frac{\gamma}{2} \sigma_z \otimes \sigma_z \right),
\]

for the interacting terms appearing in the cost Hamiltonian, and single-qubit \( R_x \) rotations:

\[
R_x(\beta) = \exp\left(-i \frac{\beta}{2} \sigma_x\right),
\]

for the transverse terms representing the mixing Hamiltonian. Here, the trainable parameters \( \gamma \) and \( \beta \) represent the strength of the interaction and the angle of the rotation, respectively.

\begin{figure}
        \centering
  \textbf{Quantum circuit for QAOA N=4}  \\ \vspace{0.2cm}
        \includegraphics[scale=0.25]{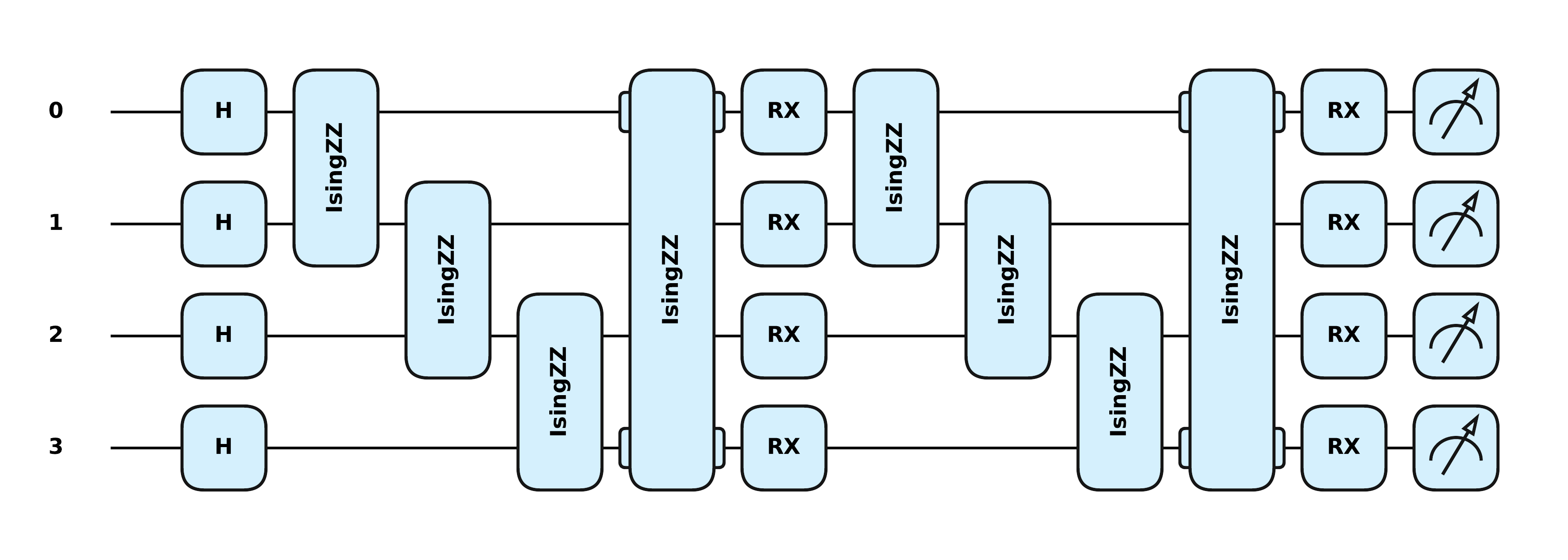}

         \caption{An example of the quantum circuit for QAOA with \( N = 4 \), implemented using the Pennylane library.}
    \label{fig:QAOA_4}
  
\end{figure}

Each QAOA layer is composed of a sequence of these gates, where all $R_x$ gates in a layer share a common trainable parameter $\beta$, and all ZZ gates share a common trainable parameter $\theta$. Importantly, all ZZ gates within a layer commute, which enables an efficient application of the interaction terms. A particular example of the quantum circuit for QAOA with \( N = 4 \) is depicted in Fig. ~\ref{fig:QAOA_4}.

\subsection{Noise Model Derived from IBM Device Calibration Data}

To simulate the realistic behavior of quantum circuits on physical hardware, we incorporated a noise model based on \texttt{ibm\_sherbrooke}'s calibration data. Using Qiskit’s Aer simulator \cite{javadiabhari2024quantumcomputingqiskit}, we utilized the backend properties of \texttt{ibm\_sherbrooke} to define noise characteristics, which were then applied during circuit execution. The calibration data for this specific device is provided in Appendix \ref{Calibration Data}, in Table \ref{Table_Calibration}. This noise model accounts for several key factors:

\begin{enumerate}
    \item \textbf{Gate Error Probability}: The likelihood of errors for each basis gate, derived from calibration data, is specific to each qubit and gate type.

    \item \textbf{Gate Duration}: Each gate’s execution time, contributing to noise exposure, is recorded from device properties.

    \item \textbf{Relaxation Time Constants}: The $T_1$ (relaxation) and $T_2$ (dephasing) times for each qubit are included, allowing for accurate modeling of qubit decoherence during computation.

    \item \textbf{Readout Error Probability}: This factor represents the probability of classical bit-flip errors in measurement results, based on qubit-specific readout performance.
\end{enumerate}

\subsubsection{Model Composition}

The noise model is implemented with the following components:

\begin{itemize}
    \item \textbf{Single-qubit gate errors}: Simulated as a combination of a single-qubit depolarizing error followed by a thermal relaxation error. The thermal relaxation parameters are obtained from each qubit’s $T_1$, $T_2$, and gate duration.

    \item \textbf{Two-qubit gate errors}: Modeled as a two-qubit depolarizing error followed by thermal relaxation errors applied to both qubits in the gate.

    \item \textbf{Measurement errors}: Single-qubit readout errors, modeled as classical bit-flip probabilities, are applied to the measured outcomes to simulate imperfections in readout fidelity.
\end{itemize}

To match the real device’s noise characteristics, the depolarizing and thermal relaxation parameters are adjusted so that the composite error rate reflects the reported gate error in the backend properties. This noise model dynamically updates based on calibration data from \texttt{ibm\_sherbrooke}, ensuring it remains accurate across different device calibrations and closely mirrors real device behavior.

\section{Numerical Results}

In this section, we report numerical experiments conducted using quantum circuits with a realistic noise model derived from an IBM quantum processor. These simulations are organized into two primary categories: (1) analysis of the fidelity of the QNG optimizer with the full Fubini-Study metric across varying circuit depths, and (2) comparison of the QNG optimizer (using both the full Fubini-Study metric and its diagonal approximation) against Vanilla Gradient Descent, both in the presence and absence of noise.

\subsection{Ground state preparation}

\begin{figure}
        \centering
 \hspace{.3cm} \textbf{Noiseless} \\
        \includegraphics[scale=0.35]{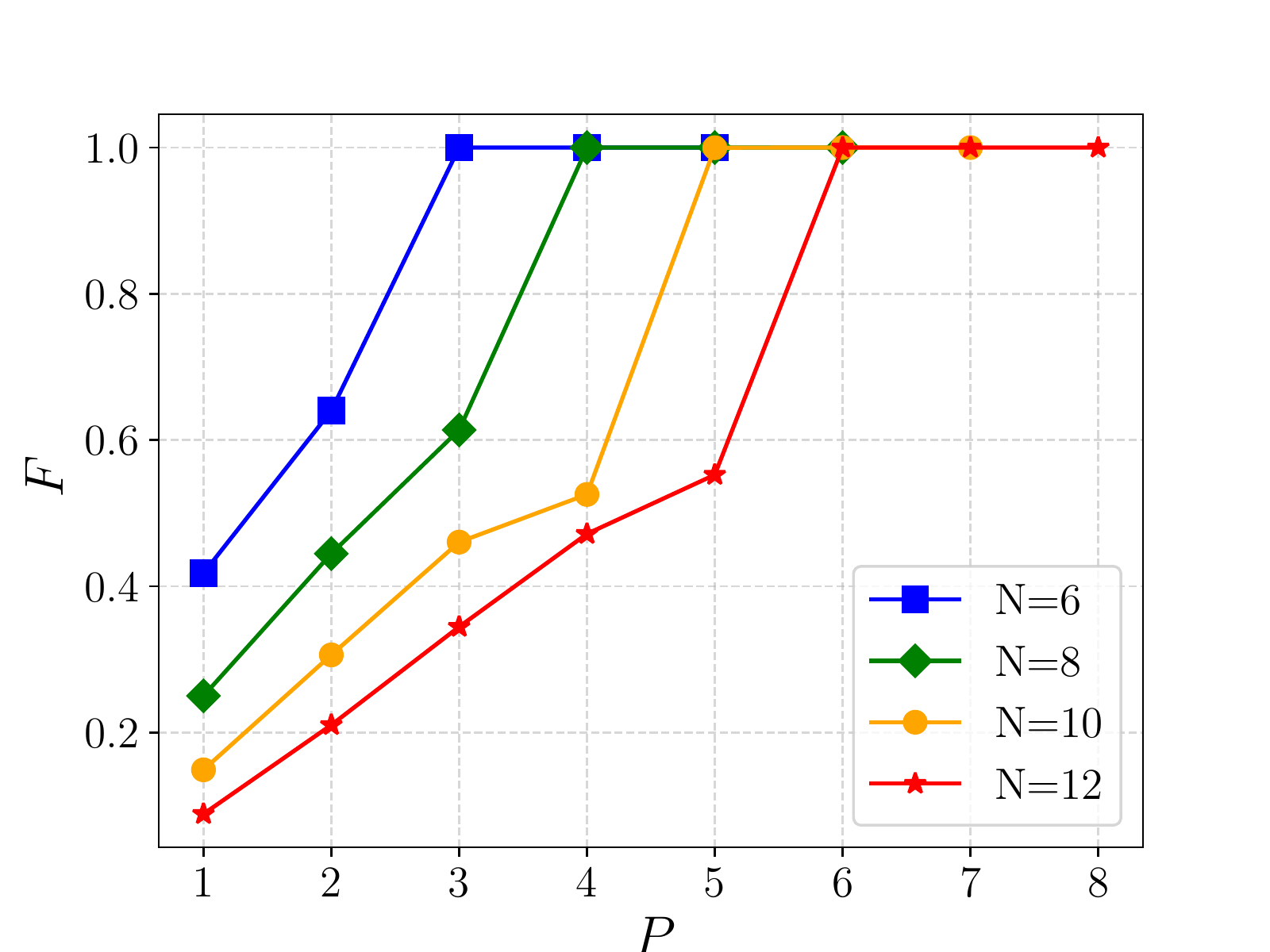}
         \caption{Fidelity growth as a function of circuit depth  \(P\) for various system sizes (\(N = [6, 8, 10, 12]\)) in the noiseless case for the quantum circuit implementation. Consistent with previous numerical studies, the highest fidelity  \(F\) is achieved at \(P = \lfloor N/2 \rfloor\).}
    \label{fig:Fidelities_vs_P_circuit_impl}
  
\end{figure}

In this experiment, we investigate how fidelity varies with the number of layers \( P \) for different system sizes, under noiseless conditions. The results are shown in  Fig.~\ref{fig:Fidelities_vs_P_circuit_impl} . In contrast to results obtained on the Rydberg atoms platform, our noiseless simulations indicate that the ground state can be achieved with an accuracy of \( 10^{-10} \) when \( P = \lfloor N/2 \rfloor \), as predicted by numerical results \cite{farhi2014quantum}. However, in the presence of noise, we were unable to reproduce the same experiment, as the current Qiskit plugin for the PennyLane library does not support outputting density matrices from the quantum circuit. Nevertheless, we expect a similar trend as in the previous implementation on the Rydberg atom platform: increasing the number of layers also increases the noise in the quantum circuit, leading to lower fidelity.

\subsection{Comparison of QNG against Vanilla gradient descent}

\begin{figure*}
      \centering
    \textbf{Noiseless accuracies} \vspace{0.2cm } \\
    \includegraphics[scale=0.25]{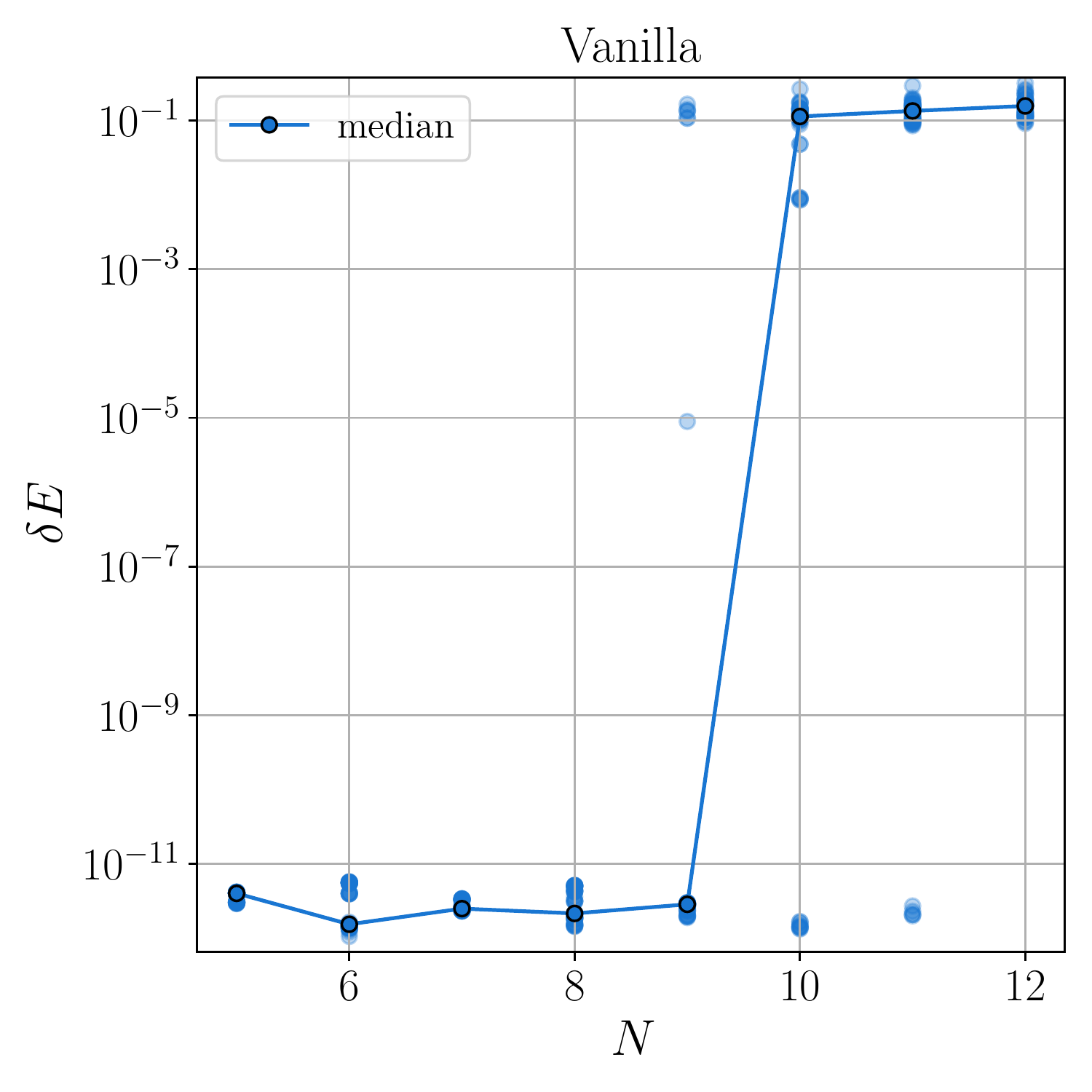}
     \includegraphics[scale=0.25]{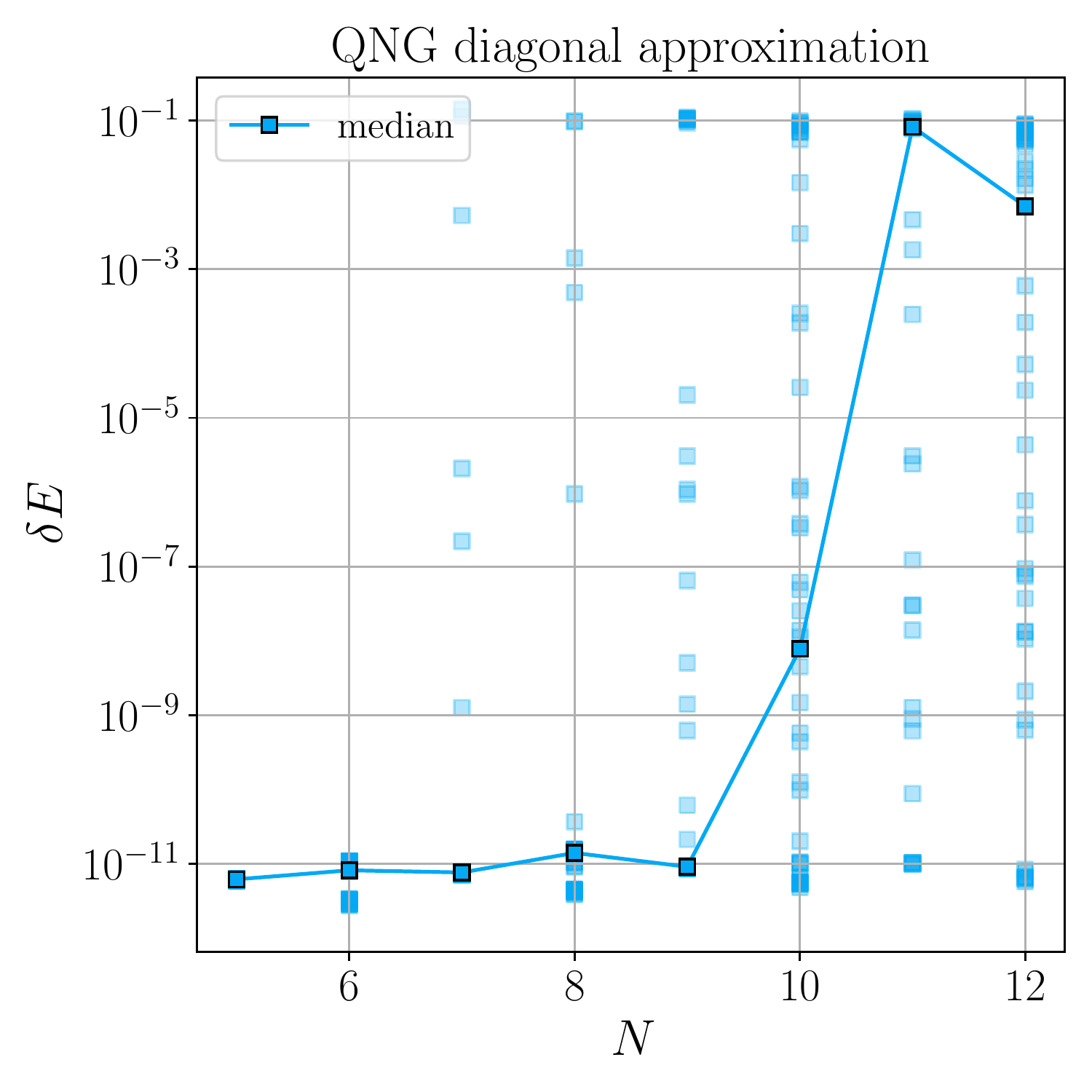}
      \includegraphics[scale=0.25]{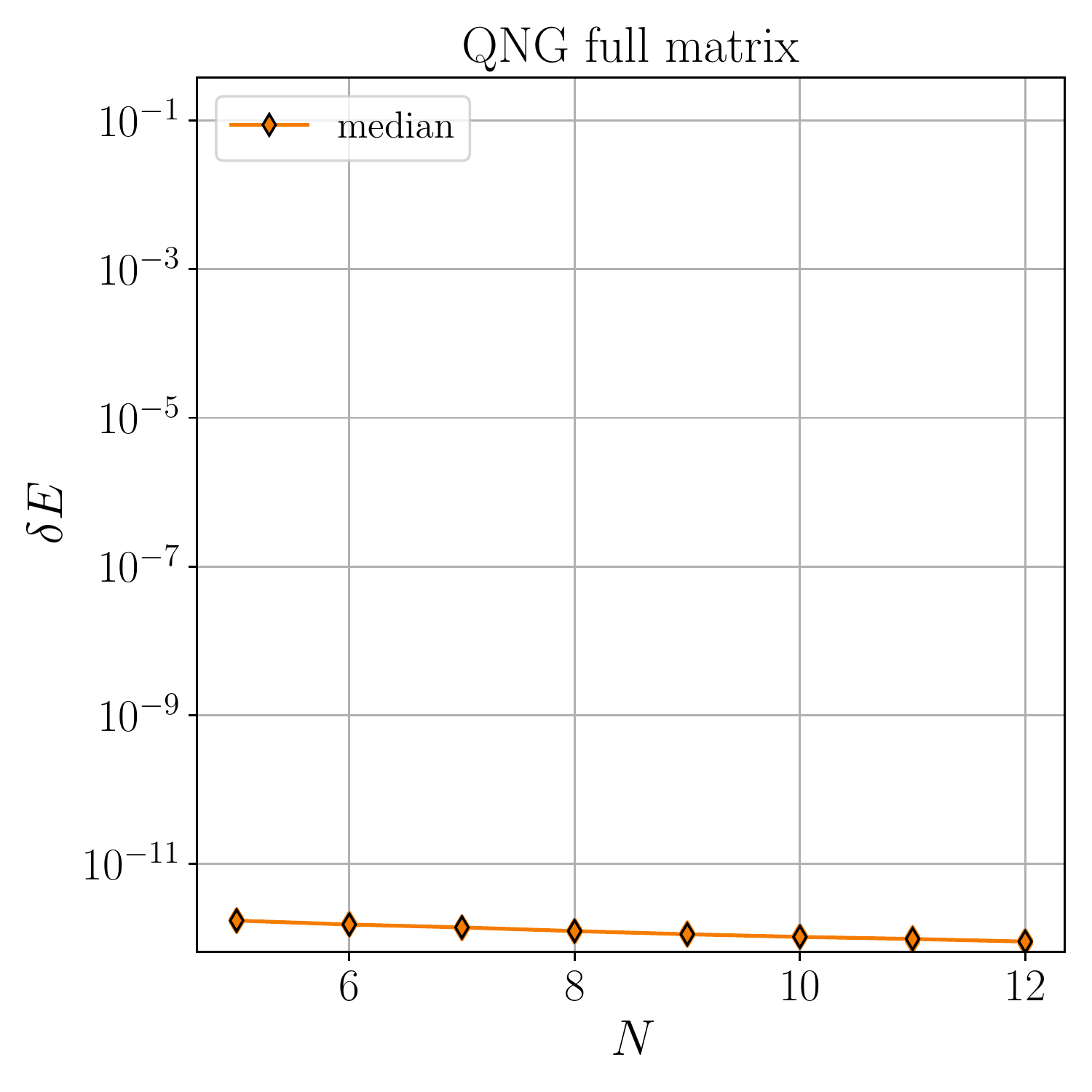}
      \includegraphics[scale=0.25]{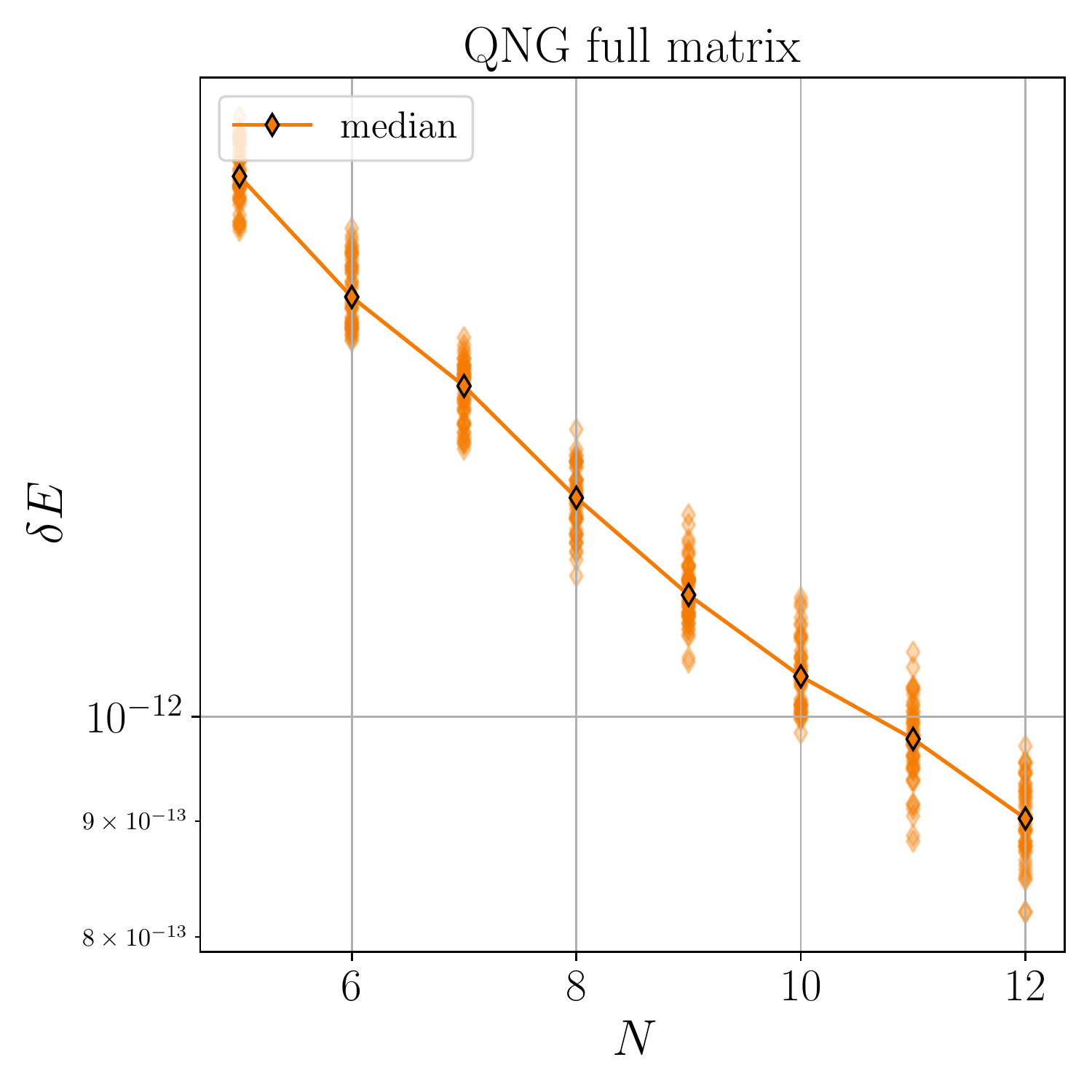}
    \caption{Comparison of noiseless accuracies for the Vanilla, QNG diagonal approximation, and QNG full matrix methods as a function of system size \( N \). Each panel displays the distribution of 50 accuracy values, measured as \( \delta E \), with the median of the accuracy represented by solid points. The rightmost plot is a zoomed-in view of the QNG full matrix accuracies.}
    \label{fig:acc_noiseless_QC}
\end{figure*}

We compared the performance of the QNG optimizer, using both the full Fubini-Study metric and its diagonal approximation, against Vanilla Gradient Descent in both noiseless (see Fig.~\ref{fig:acc_noiseless_QC}) and noisy scenarios.

\subsubsection{Noiseless Case}

In the noiseless case, we conducted simulations for system sizes \( N = [2,12] \) and ran 50 trials with different initial conditions to assess the robustness of the optimizers against random initializations. Additionally, we aimed to evaluate the extent to which the diagonal approximation allows the QNG optimizer to outperform Vanilla Gradient Descent. For the noiseless simulations, we used a learning rate of \( \eta = 0.01 \), and the maximum number of iterations for the optimization was \( 5000 \). Moreover, the convergence criterion to stop the optimization procedure, according to equation \ref{conv}, was \( \epsilon_{\text{stop}} = 10^{-12} \), while the convergence criterion to the true solution was the same as for Rydberg atoms, defined as \(\delta E_{opt} < 10^{-9} \),
where \( \delta E_{\text{opt}} \) represents the best accuracy obtained after optimization.

\begin{figure}[]
    \centering
    \textbf{Noiseless} \\ \vspace{0.2cm}
        \includegraphics[scale=0.26]{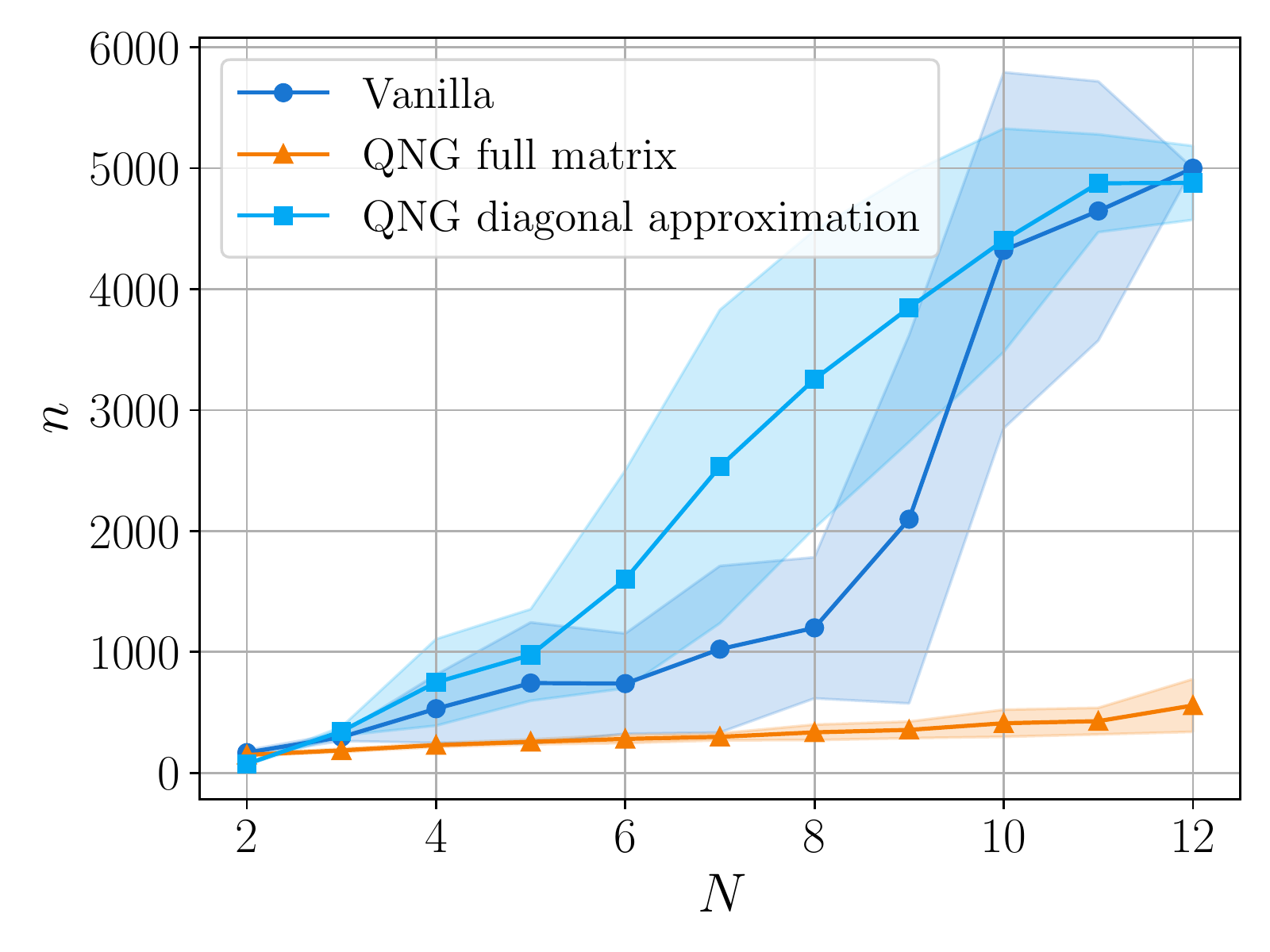}
        \includegraphics[scale=0.26]{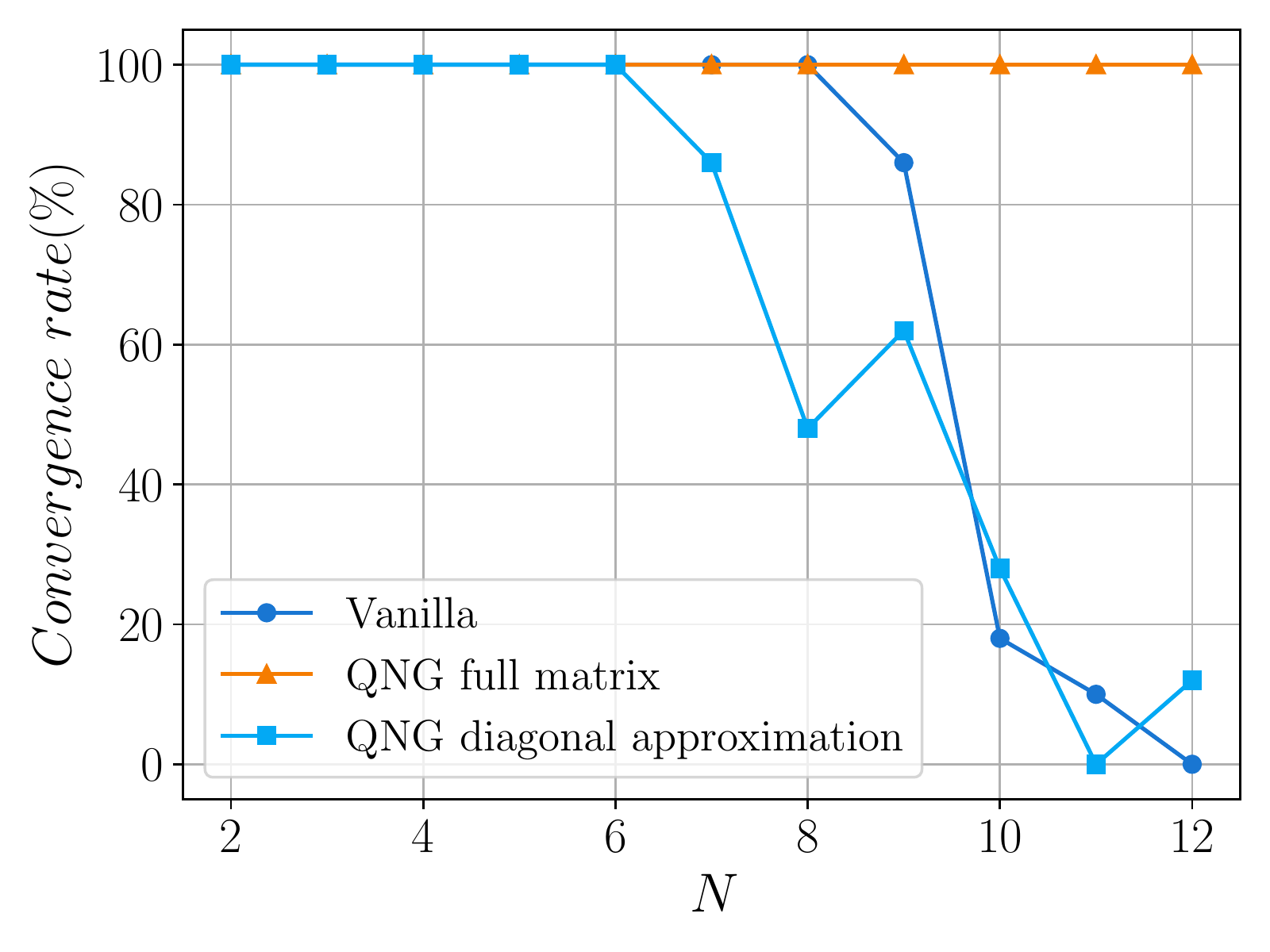}
    \caption{On the left, the average number of steps $n$ needed to reach the ground state as function of the systems size $N=[2,12]$ in the noiseless scenario. On the right, the convergence rate of QNG (full and with diagonal approximation) and Vanilla in the noiseless scenario as function of the system size $N=[2,12]$.
    }
 \label{fig:probs_ibm}
\end{figure}

As with the simulations on the Rydberg atoms platform, we plotted the average number of steps required for each optimizer to converge, along with the standard deviation across different system sizes. We also plotted the convergence rate for each optimizer across these system sizes, as shown in Fig.~\ref{fig:probs_ibm}. 

The results indicate that Quantum Natural Gradient consistently outperforms Vanilla Gradient Descent, exhibiting both faster average convergence and greater robustness to random initializations, as demonstrated by reduced variance in the number of optimization steps. Additionally, while the convergence rate of Quantum Natural Gradient remains at 100\%, the convergence rate of Vanilla declines significantly with increasing system size, highlighting the superior performance of Quantum Natural Gradient for larger problem sizes. However, for small system sizes, the performance of QNG with the diagonal approximation is inferior to that of Vanilla, exhibiting a similar decline in convergence rate. After \(N = 9\) spins, QNG begins to outperform Vanilla in terms of convergence rate, achieving a 20\% convergence rate compared to the 0\% convergence rate of Vanilla. Thus, we conclude that in this specific implementation of QAOA applied to the TFIM using quantum circuits, the diagonal approximation only becomes significant in surpassing Vanilla when the system size increases, leading to improved convergence. Meanwhile, QNG with the full Fubini-Study metric consistently achieves better results on average across noiseless experiments for all system sizes. This highlights the theoretical advantage of QNG, and its diagonal approximation, over Vanilla.

Moreover, we note that the strong robustness of QNG against random initializations and its reduced variance in optimization steps can be attributed to a distance regularization phenomenon, as mentioned earlier. This effect suggests that QNG is effectively regularizing distances within the space of quantum states. Additionally, the superior convergence rate of QNG may also indicate its ability to avoid certain local minima \cite{wierichs2020avoiding}.

Additionally, as in the numerical experiments on the Rydberg atom platform, we plotted the distribution of 50 accuracy values as a function of system size \(N\) for the different optimizers (Fig.~\ref{fig:acc_noiseless}). We observe that the QNG optimizer with the full Fubini-Study metric is the only one that maintains low accuracy with minimal variability as the system size increases, remaining below \(10^{-12}\). In contrast, the Vanilla optimizer and QNG with the diagonal approximation perform poorly as the system size increases and exhibit a wider distribution of accuracy values. 

Moreover, this reduced variance highlights the robustness of QNG against random initializations—regardless of the initial parameter configurations, QNG consistently converges close to the true solution. Regarding the comparison between the diagonal approximation and Vanilla, we observe that although both perform significantly worse than QNG with the full matrix, QNG with the diagonal approximation achieves solutions closer to the true one, with a distribution of values more shifted towards the correct solution. Thus, we conclude that even when using the diagonal approximation, better results can be obtained compared to the Vanilla optimizer.

\subsubsection{Noisy case}

For the noisy case, due to technical limitations related to the online connectivity required to obtain the noise model provided by IBM, which had to be loaded at each optimization step, we were only able to run simulations for system sizes \(N = [2, 8]\) and for 50 initial conditions. This limitation arose because the prolonged runtime of each optimization step for larger system sizes increased the likelihood of connectivity loss, leading to errors in retrieving the noise model. Nevertheless, the trend is clear enough that additional simulations are unnecessary to conclude that, in the noisy case, the algorithms with the different optimizations perform equally poor, with QNG providing no advantage. In the optimization, we used the same learning rate \( \eta \) and maximum number of iterations as in the noiseless case. However, the convergence criterion and the condition for achieving the true solution were relaxed to account for potential imperfections caused by noise. These were set to \( \epsilon_{\text{stop}} = 10^{-8} \) and \( \epsilon_{\text{true}} = 10^{-6} \), respectively.

\begin{figure}
        \centering
 \hspace{.5cm} \textbf{Noisy}  \\ \vspace{0.2cm}
        \includegraphics[scale=0.275]{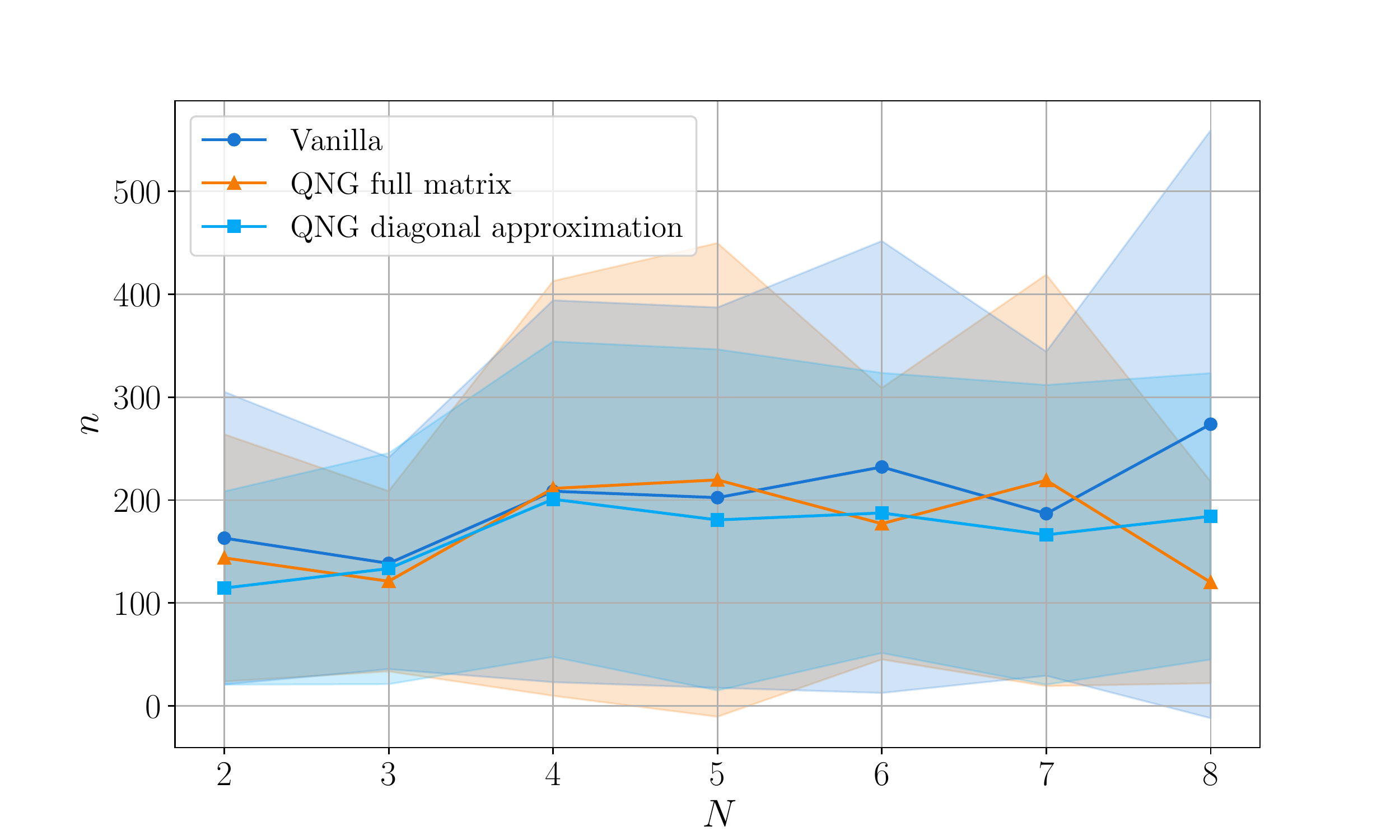}

         \caption{Average number of steps $n$ needed for convergence as function of the system's size $N=[2,8]$ in presence of noise.}
    \label{fig:mean_steps_noise}
  
\end{figure}

\begin{figure*}
      \centering
    \textbf{Noisy accuracies} \vspace{0.2cm } \\
    \includegraphics[scale=0.3]{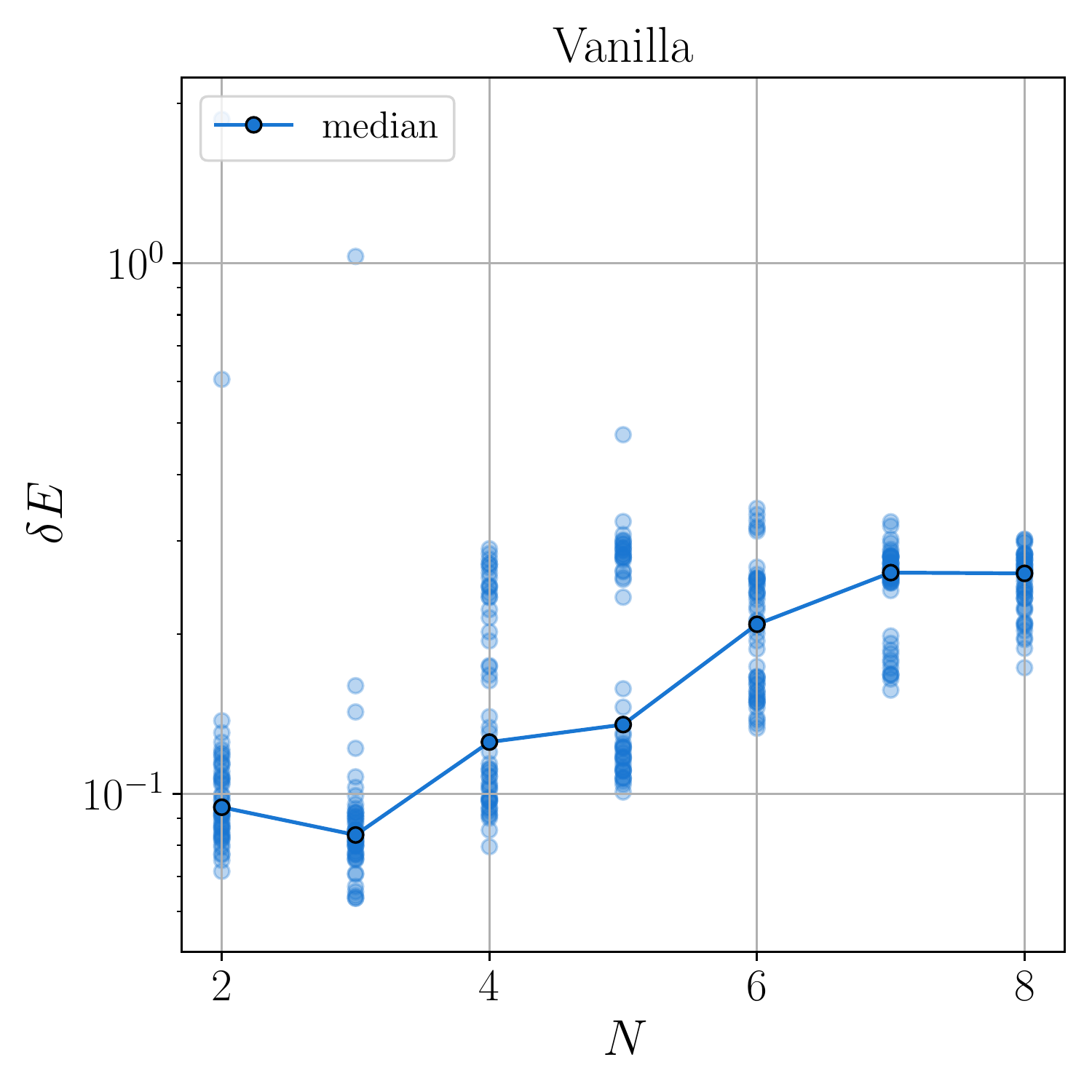}
     \includegraphics[scale=0.3]{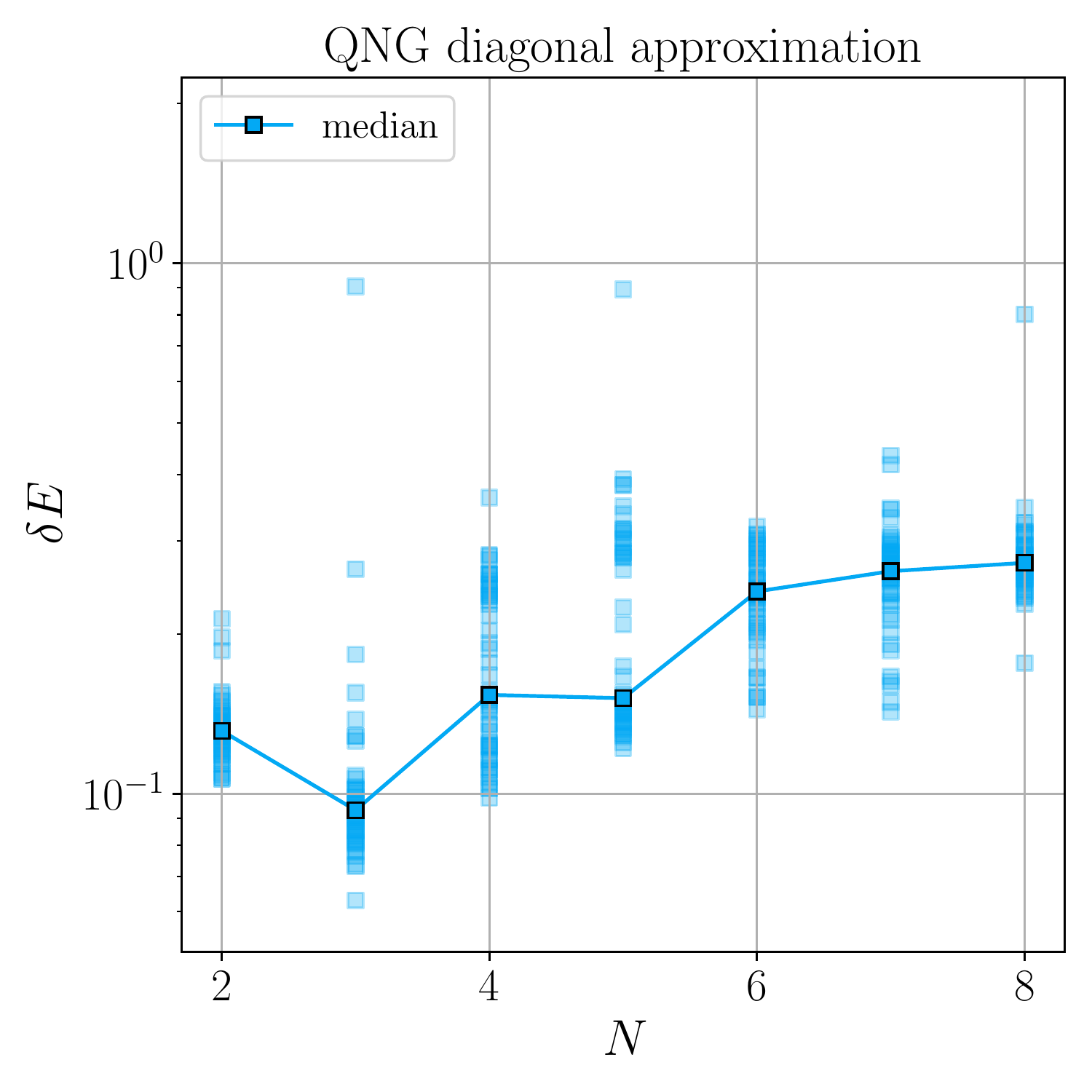}
      \includegraphics[scale=0.3]{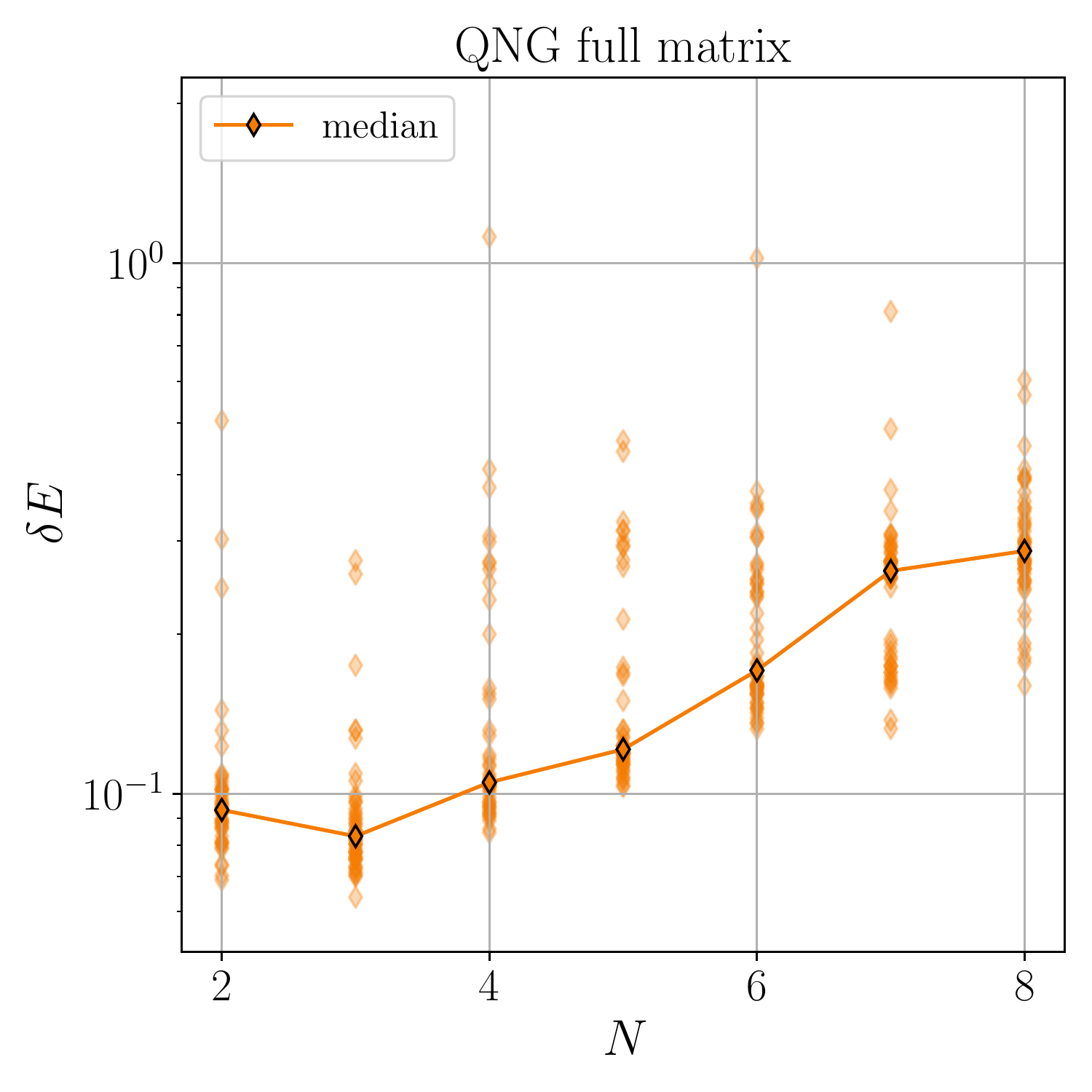}
    \caption{Comparison of noisy accuracies for the Vanilla, QNG diagonal approximation, and QNG full matrix methods as a function of system size \( N \). Each panel displays the distribution of 50 accuracy values, measured as \( \delta E \), with the median of the accuracy represented by solid points.}
    \label{fig:acc_noisy}
\end{figure*}

In these experiments, due to the inability of the algorithms to reach the true solution because of noise, we decided not to plot the convergence rate but rather the accuracy (see \ref{eq:accuaracy}) obtained from the 50 runs for each optimizer. We can observe the plot for the average number of steps for the different optimizers in Fig.~\ref{fig:mean_steps_noise}. Additionally, the accuracy plots for the different optimizers are shown in Fig.~\ref{fig:acc_noisy}.

From the accuracy plot, it is evident that none of the optimizers is able to prepare the true ground state of the TFIM using this realistic noise model, as indicated by the high values of accuracy. As previously mentioned, this may arise because this problem is not well-suited for noisy digital quantum computers, resulting in gate decomposition into native gates that leads to overhead. This, combined with high noise levels, yields very poor outcomes for the quantum algorithm. Furthermore, the trend in the average number of steps is relatively similar across the three optimizers. Thus, we can conclude that for QAOA applied to the TFIM using quantum circuits with a realistic noise model from an IBM quantum device, QNG offers no advantage over Vanilla Gradient Descent, as both optimizers yield unsatisfactory results.

\section{Conclusions and outlook}\label{Conclusions}

This work evaluated the performance of the Quantum Natural Gradient (QNG) optimizer within the Quantum Approximate Optimization Algorithm (QAOA), benchmarking its efficacy on two prominent quantum computing platforms: Rydberg atoms and superconducting circuits. The results provide significant insights into the potential of the QNG optimizer to enhance the optimization of variational quantum algorithms, even in the presence of noise, such as that encountered in noisy intermediate-scale quantum (NISQ) devices.

Across both platforms, QNG consistently outperformed Vanilla Gradient Descent, achieving faster convergence and greater robustness to random initializations. This suggests that QNG effectively regularizes distances within the space of quantum states. Furthermore, QNG demonstrated a higher convergence rate to the true solution, successfully avoiding certain local minima. On the Rydberg atoms platform, the diagonal approximation of the Quantum Fisher Information (QFI) offered a computationally efficient alternative without significantly compromising performance, making it a viable choice for larger systems.

Rydberg atom platforms demonstrated superior performance compared to superconducting circuits. The analog implementation on Rydberg systems eliminates the need for local qubit addressing, a common experimental challenge in digital platforms. Furthermore, Rydberg platforms exhibited remarkable robustness against noise, enabling ground state preparation with high fidelity under realistic conditions. In contrast, superconducting circuits struggled under noise, with decoherence effects so severe that ground state preparation became unattainable. This limitation was further exacerbated by the gate decomposition overhead required to implement QAOA on digital hardware, which increases circuit depth and amplifies noise effects.

Simulations on superconducting platforms highlighted their sensitivity to noise. Despite QNG’s theoretical advantages, the combined effects of gate decomposition and noise rendered the algorithm ineffective for state preparation in this setting.

While the Quantum Natural Gradient appears to be a very promising method for enhancing variational algorithms, it is important to consider the cost associated with computing the full Quantum Fisher Information Matrix, which becomes impractical for a large number of parameters. Future research should focus on finding methods to reduce the scaling of the computation of the full QFIM. This could include restricting the approach to specific variational architectures with favorable scaling or developing improved protocols to compute it using fewer quantum resources. 

Additionally, new approximations to the QFIM should be investigated, particularly for efficiently computing the off-block-diagonal terms. A natural next step for this work would be to incorporate shot noise into the numerical simulations and study how the inverse QFIM behaves under these conditions, as well as how this affects the performance of QNG. Another potential research direction would be to explore ways to derive a simpler expression for the QFIM in the presence of noise, beyond its current formulation in terms of the diagonalization of the density matrix.

In summary, the QNG optimizer is a highly advantageous method for enhancing the optimization of variational algorithms. It enables faster average convergence, greater robustness against random parameter initializations, and an improved ability to avoid certain local minima. Furthermore, our results demonstrate that QNG exhibits robustness against realistic noise during optimization. Thus, we firmly believe that incorporating QNG for the optimization of variational quantum algorithms is highly promising, provided that significant progress is made in reducing the computational cost of gradient evaluation and QFIM estimation on real hardware.

Finally, we want to emphasize that, despite the strong potential of QNG---offering more stable convergence and improved resilience to noise---these benefits come with significant computational costs. In particular, evaluating the QFIM, whether exactly or approximately, introduces substantial overhead in both classical post-processing and quantum measurement resources. As a result, the practical use of QNG is currently limited to small-scale simulations or hypothetical future scenarios where these costs may become more manageable.

\section{Acknowledgments}
This work was supported by the project PID2023-152724NA-I00, with funding from MCIU/AEI/10.13039/501100011033 and FSE+, the Severo Ochoa Grant CEX2023-001292-S, Generalitat Valenciana grant CIPROM/2022/66, the Ministry of Economic Affairs and Digital Transformation of the Spanish Government through the QUANTUM ENIA project call - QUANTUM SPAIN project, and by the European Union through the Recovery, Transformation and Resilience Plan - NextGenerationEU within the framework of the Digital Spain 2026 Agenda, and by the CSIC Interdisciplinary Thematic Platform (PTI+) on Quantum Technologies (PTI-QTEP+). This project has also received funding from the European Union’s Horizon 2020 research and innovation program under grant agreement CaLIGOLA MSCA-2021-SE-01-101086123. RGL is funded by grant CIACIF/2021/136 from Generalitat Valenciana.
The work was also partially supported by INFN through the project QUANTUM.
\bibliography{biblio}

\newpage
\onecolumngrid
\appendix
\section{Calibration data from the IBM device \texttt{ibm\_sherbrooke}}\label{Calibration Data}

In this Appendix, we present the calibration data from the IBM device \texttt{ibm\_sherbrooke}, which were used as parameters for our noise model during circuit execution. The calibration data is shown in Table \ref{Table_Calibration}.

\begin{table}[h!]
\resizebox{\textwidth}{!}{%
\begin{tabular}{@{}llllllllllll@{}}
\toprule
Qubit & T1 (us) & T2 (us) & Prob meas0 prep1 & Prob meas1 prep0 & Readout assignment error & ID error  & RZ error & SX error  & X error   & ECR error                          & Gate time (ns)                     \\ \midrule
0     & 351.63  & 252.89  & 0.26             & 0.23333          & 0.24667                  & 0.0001205 & 0        & 0.0001205 & 0.0001205 &                     &    \\
1     & 391.24  & 383.13  & 0.0172           & 0.0228           & 0.02                     & 0.0001565 & 0        & 0.0001565 & 0.0001565 & ecr1\_2:0.0049;ecr1\_0:0.0064      & ecr1\_2:533.333;ecr1\_0:533.333    \\
2     & 268.69  & 170.39  & 0.0194           & 0.0338           & 0.0266                   & 0.0002395 & 0        & 0.0002395 & 0.0002395 &      &     \\
3     & 281.24  & 259.83  & 0.53667          & 0.35             & 0.44333                  & 0.0002122 & 0        & 0.0002122 & 0.0002122 & ecr3\_2:0.00619    & ecr3\_2:533.333    \\
4     & 410.16  & 440.13  & 0.65667          & 0.2              & 0.42833                  & 0.0001117 & 0        & 0.0001117 & 0.0001117 & ecr4\_3:0.00444    & ecr4\_3:533.333    \\
5     & 263.77  & 290.03  & 0.58             & 0.30333          & 0.44167                  & 0.0002235 & 0        & 0.0002235 & 0.0002235 & ecr5\_4:0.00742    & ecr5\_4:533.333   \\
6     & 320.37  & 263.67  & 0.56             & 0.28333          & 0.42167                  & 0.0002429 & 0        & 0.0002429 & 0.0002429 & ecr6\_5:0.00575    & ecr6\_5:533.333    \\
7     & 342.18  & 287.59  & 0.27333          & 0.46             & 0.36667                  & 0.0034641 & 0        & 0.0034641 & 0.0034641 & ecr7\_6:0.00762;ecr7\_8:0.01442    & ecr7\_6:533.333;ecr7\_8:533.333    \\
8     & 246.37  & 247.17  & 0.49             & 0.30667          & 0.39833                  & 0.0001891 & 0        & 0.0001891 & 0.0001891 & ecr8\_9:0.0054    & ecr8\_9:533.333    \\
9     & 397.17  & 69.63   & 0.59333          & 0.31             & 0.45167                  & 0.0006543 & 0        & 0.0006543 & 0.0006543 &    &    \\
10    & 272.59  & 192.26  & 0.0134           & 0.0044           & 0.0089                   & 0.0002901 & 0        & 0.0002901 & 0.0002901 & ecr10\_11:0.00676;ecr10\_9:0.01087 & ecr10\_11:533.333;ecr10\_9:533.333 \\
11    & 250.45  & 93.71   & 0.45667          & 0.41333          & 0.43500                  & 0.0001855 & 0        & 0.0001855 & 0.0001855 &                &                  \\ \bottomrule
\end{tabular}%
}
\caption{Calibration data from the IBM quantum device \texttt{ibm\_sherbrooke}, which was used to construct the realistic noise model. The ECR gate, used as a basis gate in IBM quantum processors, is a maximally entangling gate and is equivalent to a CNOT gate up to single-qubit pre-rotations. The numbers indicate the qubits to which the gate is applied and the order in which it is applied. Note that blank spaces for the ECR gate in some rows indicate that the gate is only available in the specific order given by the table.}
\label{Table_Calibration}
\end{table}

\end{document}